\documentclass[12pt]{article}
\usepackage{jcappub}
\usepackage{adjustbox,multirow, longtable}
\usepackage{cancel,enumitem,dsfont,extarrows}
\usepackage{mathtools,amssymb,amsthm,amsmath,amsfonts}
\usepackage{xcolor,graphicx,subfigure}
\usepackage[utf8]{inputenc}
\usepackage[normalem]{ulem}
\allowdisplaybreaks

\newcommand{\ignore}[1]{}

\definecolor{darkgreen}{RGB}{50,150,0}

\arxivnumber{2502.16082}
\begin{document}
\title{Calculating the EFT likelihood via saddle-point expansion}
\author[a]{Ji-Yuan Ke,}
\author[a]{Yun Wang}
\author[a,b]{and Ping He}
\affiliation[a]{Center for Theoretical Physics and College of Physics, Jilin University, Changchun 130012, China}
\affiliation[b]{Center for High Energy Physics, Peking University, Beijing 100871, China}
\emailAdd{kejy22@mails.jlu.edu.cn, yunw@jlu.edu.cn, hep@jlu.edu.cn}

% -----------------------------------------------------------------------------------------------
\abstract
{In this paper, we extend the functional approach for calculating the EFT likelihood by applying the saddle-point expansion. We demonstrate that, after suitable reformulation, the likelihood expression is consistent with the path integral required to be computed in the theory of false vacuum decay. In contrast to the saddle-point approximation, the application of the saddle-point expansion necessitates more nuanced considerations, particularly concerning the treatment of the negative eigenvalues of the second derivative of the action at the saddle point. We illustrate that a similar issue arises in the likelihood calculation, which requires approximating the original integral contour through the combination of the steepest descent contours in the field space. As a concrete example, we focus on calculating the EFT likelihood under a Gaussian distribution and propose a general procedure for computing the likelihood using the saddle-point expansion method for arbitrary partition functions. Precise computation of the likelihood will benefit Bayesian forward modeling, thereby enabling more reliable theoretical predictions.}

\maketitle
%%%%%%%%%%%%%%%%%%%%%%%%%%%%%%%%%%%%%%%%%%%%%%%%%%

% %%%%%%%%%%%%%%%%%%%%%%%%%%%%%%%%%%%%%%%%%%%%%%%%%%
\section{Introduction}

The large-scale structure (LSS) of the universe encodes valuable information about galaxy formation, structure growth, and the physics of dark matter and dark energy. To extract this information, we need to fit our theoretical predictions to the measurements from observations. Conventionally, the galaxies are regarded as the tracers of the dark matter distribution, and their $n$-point correlation function is used to compare with data to provide constraints on cosmological parameters (see Ref.~\cite{Desjacques:2016bnm} for a comprehensive review). 

In recent years, Bayesian forward modeling \cite{Jasche:2012kq, Wang:2014hia, Jasche:2018oym, Lavaux:2019fjr, Kitaura:2019ber, Bos:2018rpw} has emerged as another powerful approach for studying such problems, which aims to directly predict the present-day galaxy distribution given different initial conditions, galaxy bias models, and cosmological parameters, then iteratively compare the resulting galaxy distribution with observations until convergence. The key advantage of this method is that it fully exploits all available information without the need for summary statistics. Consequently, Bayesian forward modeling has been widely applied to field-level analyses of galaxy clustering \cite{Nguyen:2024yth, Kostic:2022vok, Stadler:2024fui, Ibanez:2023vxb, Stadler:2023hea, Lazeyras:2021dar, Nguyen:2020hxe, Philcox:2020srd, Elsner:2019rql, Schmidt:2020viy, Barreira:2021ukk}.

A necessary basis in applying the Bayesian method is to calculate the conditional likelihood $\mathcal{P}[\delta_g|\delta]$, which is the functional probability of observing the galaxy overdensity $\delta_g$ given the matter overdensity $\delta$ (see Refs.~\cite{Cabass:2019lqx, Schmidt:2018bkr} for more details). To achieve this, we need to factorize $\mathcal{P}[\delta_g|\delta]$ into two components, $\mathcal{P}[\delta]$ and $\mathcal{P}[\delta_g,\delta]$, which are referred to as the matter likelihood and joint likelihood, respectively. Thanks to the recently developed path integral approach to the LSS \cite{Carroll:2013oxa, Cabass:2019lqx}, both likelihoods can be naturally expressed as the functional Fourier transform of the partition functions $Z[\delta]$ and $Z[\delta_g,\delta]$. Meanwhile, the Effective Field Theory of LSS (EFTofLSS) \cite{Baumann:2010tm, Carrasco:2012cv, Carroll:2013oxa, Porto:2013qua, Carrasco:2013mua, Konstandin:2019bay, Senatore:2014via, Senatore:2014eva, Senatore:2014vja} postulates that the contribution of short-wavelength (small-scale) fluctuations to the long-wavelength (large-scale) universe can be considered as a sequence of non-ideal fluid effects which can be absorbed into the corresponding terms of the partition functions, such as the pressure perturbation and the viscosity. It is therefore important to establish a framework to derive the EFT likelihood (hereinafter referred to as likelihood) from a general partition function. However, given general $Z[\delta]$ and $Z[\delta_g,\delta]$, it is usually unfeasible to evaluate $\mathcal{P}[\delta]$ and $\mathcal{P}[\delta_g,\delta]$ by direct calculation because of the complicated functional integral formula. In \cite{Cabass:2019lqx}, a functional approach for computing the conditional likelihood through the saddle-point approximation is proposed, yielding results that are nearly consistent with the traditional EFT estimate, except for additional higher-order terms \cite{Schmidt:2018bkr}. Furthermore, these higher-order terms are identified as precise corrections to the conditional likelihood expression.

The present paper aims to calculate the likelihoods more precisely. Since the saddle-point contribution in the exponent provides a good approximation to the full likelihood, it is natural to extend this approach to higher orders around the saddle points, i.e., via the saddle-point expansion. This semiclassical method was first proposed in quantum mechanics and quantum field theory to calculate the decay rates and leads to reliable results that are consistent with the quantum-mechanical estimates of tunneling rates \cite{Coleman:1977py,Callan:1977pt}. More recently, the saddle-point expansion method has also been applied to Yang-Mills theory, leading to the CP conservation in the strong interactions \cite{Ai:2020ptm,Ai:2024cnp}. In this paper, we expect to apply this method within the path-integral formulation of LSS to achieve a more accurate prediction of the conditional likelihood. 

In the process of applying the saddle-point expansion, there is a notorious problem corresponding to the negative eigenvalues of the second derivative of the classical action at the saddle point. When we substitute the saddle-point solution into the integral, this mode will change the sign on the exponent of the path integral, thus making the integral ill-defined. The resolution to the problem, is to apply the Picard-Lefschetz theory \cite{Witten:2010cx,Ai:2019fri}, deforming the integral contour into the combination of the steepest descent contours which connect different saddle points, and terminate at the convergent regions of the integral. In this way, the imaginary components of different ``sub-contours" (often referred to as the Lefschetz thimbles) will cancel each other out to give the real result \cite{Andreassen:2016cvx}. In this paper, we will show that we also encounter negative modes, and even complex modes in the calculation of the likelihoods, and we can make use of the steepest descent contour method to guarantee the result real. This is the kernel of our work because the result has the physical meaning of probability.

One of our main results is that the saddle-point expansion method used in the calculation of decay rates can be perfectly applied to the calculation of the likelihoods. What we need to do is simply reformulate the likelihoods into the appropriate form. Take the case of the joint likelihood as an example, we can write
\begin{equation}
   \mathcal{P}[\delta_g,\delta] = \int \mathcal{D} \boldsymbol{\phi}_g \, e^{-S_g[\boldsymbol{\phi}_g]} \, ,
\end{equation}
where $\boldsymbol{\phi}_g$ is the field that absorbs all the integral parameters and $S_g[\boldsymbol{\phi}_g]$ is the ``modified action". Then we can use the same arguments as in quantum mechanics and quantum field theory to solve the expression for the likelihood analytically. At the same time, a problem that is not the same as the situation in quantum field theory is that the action is often not real because its expression often contains the imaginary unit $i$. However, in Sec.~\ref{sec:example} we will prove that although $S_g[\boldsymbol{\phi}_g]$ and the eigenvalues of $S_g''[\bar{\boldsymbol{\phi}}_g]$ may be complex, we can still guarantee the integral result given by the sum of different steepest descent contours is real.

Also, we will illustrate how to apply the saddle-point expansion method to compute the likelihood for a given $S_g[\boldsymbol{\phi}_g]$ by a concrete example, i.e. the EFT likelihood with a Gaussian distribution. We will provide a detailed introduction on how to use the gradient flow equation to find the steepest descent contours in the field space. We emphasize here that, although we have only studied a specific model, our discussion of this method and the treatment of the negative (complex) modes do not depend on the form of the action. Our approach is general, and can be applied to more complicated situations.

The organization of this paper is as follows. In Sec.~\ref{sec:qmandqft}, we review the saddle-point expansion method in quantum mechanics and quantum field theory to calculate the decay rates. The analogy between quantum field theory and LSS path-integral case will be drawn in Sec.~\ref{sec:saddleofLSS}. Sec.~\ref{sec:example} contains an example of applying this method to calculate the conditional likelihood. Then we will draw our conclusion and discuss future directions in Sec.~\ref{sec:conclusion}. To complete our work, we provide a direct calculation illustrating the cancellation of the imaginary contributions arising from different steepest descent contours in Appendix~\ref{sec:cancellation}, and we also review how to calculate the contribution of the saddle points to the likelihood in Appendix~\ref{sec:saddle-point-contribution}. We present detailed discussions on how to solve the eigenvalues of $S''_{g}[\bar {\boldsymbol{\phi}}_g]$ in Appendix~\ref{sec:eigenvalues}.

Throughout this paper, the notations and conventions used in the path-integral approach to the LSS mainly come from \cite{Cabass:2019lqx}.

% ----------------------------------------------------------------------------------------------------------
\section{Saddle-point expansion in quantum mechanics and quantum field theory}
\label{sec:qmandqft}

In this section, we will review the application of the saddle-point expansion in quantum mechanics and quantum field theory, inspired by the calculation of decay rates \cite{Coleman:1977py,Callan:1977pt,Andreassen:2016cvx,Ai:2019fri}. We will see that the existence of the negative modes of the second derivative of the action at the saddle points compels us to complexify the integral path and deform the integral contour. We need to approximate the integral contour by the sum of the steepest descent contours connecting different saddle points, which is referred to as the combination of different Lefschetz thimbles in the Picard-Lefschetz theory \cite{Witten:2010cx, Witten:2010zr}. Our review contains only the most central parts of the saddle-point expansion method, while more detailed and rigorous reviews can be found in \cite{Andreassen:2016cvx, Ai:2019fri}.

\subsection{Quantum-mechanical case}
\label{qmcase}
\begin{figure}
    \centering
    \includegraphics[width=0.7\linewidth]{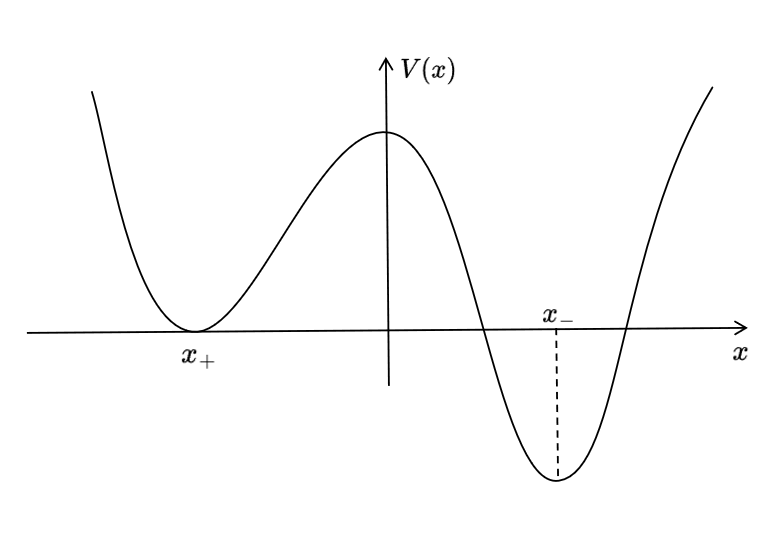}
    \caption{The classical double-well potential that is often used in false vacuum decay theory. The false vacuum state and the true vacuum state are labeled by $x_+$ and $x_-$.}
    \label{fig:double-well}
\end{figure}

We start from the quantum-mechanical case. As originally proposed by Coleman and Callan \cite{Coleman:1977py,Callan:1977pt}, in the computation of decay rates, one should consider the Euclidean-space path integral
\begin{equation}
    \mathcal{Z} \equiv \left\langle x_+ | e^{-H \mathcal{T}} | x_+\right\rangle = \int\mathcal{D}x(\tau) \, e^{-S_{E}[x(\tau)]}\, , \label{qmpath}
\end{equation}
where $S_{E}[x]$ is the classical Euclidean action, related to the Euclidean Lagrangian via $S_{E}[x] = \int {\rm d} \tau \mathcal{L}_{E} =\int {\rm d}\tau [\frac{1}{2}(\frac{{\rm d}x}{{\rm d}\tau})^2 +V(x)]$. Here, $\tau$ represents the Euclidean time coordinate, and $\mathcal{T}$ denotes the Euclidean time interval after the Wick rotation. For convenience, the potential is often taken as a double-well formula with different depths, as shown in Fig.~\ref{fig:double-well}. We denote the false vacuum as $x_+$, and the true vacuum as $x_-$. This integral needs to be calculated along all trajectories that have the boundary condition $x(-\mathcal{T}/2) = x(\mathcal{T}/2) = x_+$.

% ----------------------------------------------------------------------------------------------------------

The first step in implementing the saddle-point expansion is to decompose the trajectories $x(\tau)$ into two components: the classical paths and the quantum fluctuations
\begin{equation}
    x(\tau) = \bar{x}(\tau)\, + \,\tilde{x}(\tau)\,,
\end{equation}
where $\bar{x}$ is the stationary point of the action, satisfying $(\delta S[{x}]/\delta x)|_{x=\bar{x}} = 0$, equivalent to the classical equation of motion (EOM)
\begin{equation}
    \frac{{\rm d}^2 x(\tau)}{{\rm d} \tau^2} - V'(x(\tau)) = 0\,, \label{qmeom}
\end{equation}
where the prime denotes differentiation with respect to $x$. It is evident that the EOM in Euclidean space, compared to its counterpart in Minkowski spacetime, is equivalent to a sign flip of the potential term. Consequently, in the context of general false vacuum decay theory, this EOM Eq. \eqref{qmeom} typically admits three solutions: (1) The trivial false-vacuum static solution $x_{F}(\tau) \equiv x_+$. (2) The bounces solution $x_{B_n}(\tau)$\footnote{In quantum field theory, since the bounces solution exists only in Euclidean space rather than Minkowski spacetime, it is also known as the instanton.}, which refers to a trajectory that starts with zero velocity, bounces out of the potential well $n$ times, and eventually returns to the origin again, as illustrated in \cite{Coleman:1977py}. (3) The last one is the shot solution $x_{S}(\tau)$, described in \cite{Andreassen:2016cff, Andreassen:2016cvx}, which quickly reaches the other side of the potential well at $V(x) = 0$, where it stays for the majority of the time, before ultimately returning to the origin. These three solutions are vividly illustrated in Fig.~5 of \cite{Andreassen:2016cvx}. Notably, the bounces solution does not correspond to a local minimum of the action but rather to a saddle point. To emphasize the essence of the saddle-point expansion, we omit the detailed summation over the multi-bounces and subsequently denote it succinctly as $B$ or bounce (the details for the summation process can be found in, e.g., \cite{Ai:2019fri}).

The classical part will dominate the path integral (if we recover $\hbar$ in this theory, then the coordinate decomposition should be $ x = \bar{x}\, + \,\hbar^{1/2}\tilde{x}$, see for example in \cite{Garbrecht:2015oea}), thus we can expand the action around all the stationary points $x = \bar{x}_n$ ($n = FV$, $B$, and $S$). Then the path integral can be formulated as
% --------------------------------------------------------------------------------------------------------------
\begin{equation}
\int\mathcal{D} x(\tau) \,  e^{-S_{E}[x(\tau)]}\,  \approx \sum_{n=FV, B, S} \int \mathcal{D}\tilde{x}_{n}(\tau) \,  e^{-S_{E}[\bar{x}_n] - \frac{1}{2}S''_{E}[\bar{x}_n] \tilde{x}_n^2},
\label{qmsaddlepoint}
\end{equation}
where on the right-hand side (RHS) the linear term vanishes because $\bar{x}_n$ are the solutions of the EOM. The issue emerges when we proceed to the subsequent step: to evaluate the RHS of Eq.~\eqref{qmsaddlepoint}, it is necessary to determine the eigenvalues of $S_{E}''[\bar{x}]$. However, for the single bounce solution of Eq.~\eqref{qmeom} there is always a negative mode, making the path-integral ill-defined \cite{Callan:1977pt}. There is also a zero mode proportion to $dx/d\tau$. To handle the zero mode, we can convert it into a collective coordinate \cite{Coleman:1978ae, Andreassen:2016cvx, Bhattacharya:2024chz}. However, the treatment of the negative mode involves several subtle aspects, which will be discussed in greater detail below.

In fact, the resolution of this problem requires the application of the steepest descent contour method. The main strategy consists of the following steps: (1) We complexify the action and identify all the complex saddle points, denoted by $s_1,s_2\cdots,s_n$. (2) For each $s_i$ in the complex plane, we determine a corresponding steepest descent contour $C_i$\footnote{We emphasize that the choice of the steepest descent contour is not unique. In fact, all contours that render the integral finite are called integration cycles. These cycles and their equivalence relations form a homology group. What is required is the contour that is homologous to the original one, as illustrated in \cite{Ai:2019fri}.} (also known as the Lefschetz thimble \cite{Witten:2010cx}). Concretely, each $C_i$ is defined simply by moving away from $s_i$ in the direction that increases the real part of $S$ as quickly as possible. (3) Following the above spirit, the Lefshcetz thimble corresponding to each $s_i$ either terminates at the convergent regions of the integral or another saddle point. Therefore, each thimble provides a convergent path integral along the complex integration contour in field space.  (4) With these thimbles in field space, we can approximate the integral contour as the combination of different thimbles fulfilling the conditions, each thimble corresponds to
\begin{align}
    \mathcal{J}_i =\int_{C_i} {\rm d}z \,  e^{-S_{E}[z]} \approx \int_{C_i} {\rm d}z \,  e^{-S_{E}[s_i] - \frac{1}{2}S''_{E}[s_i] (z-s_i)^2 +\cdots} 
     \sim \sqrt{\frac{2\pi}{S''_{E}[s_i]}}\, e^{-S_{E}[s_i]} \equiv \mathcal{I}_i\,, \label{approxma}
\end{align}
where $C_i$ distinguishes different Lefschetz thimbles, and on the RHS, we have written the parameter of $S$ as $z$ to illustrate the action has been complexified. Then the full path integral can be expressed as follows
\begin{equation}
    \mathcal{Z} = \sum_i\mathcal{J}_i \sim \sum_i \mathcal{I}_i\,. \label{total}
\end{equation}
where ``$\sim$" represents other corrections are exponentially small. Notably, as demonstrated in the third step of Eq.~\eqref{total}, the saddle-point expansion method effectively approximates the path integral by considering the contributions around the saddle points. Moreover, the existence of the negative modes does not indicate a flaw in the theory, but rather a consequence of incorrectly applying the steepest descent method.

\subsubsection*{Example}

\begin{figure}
    \centering
    \includegraphics[width=0.7\linewidth]{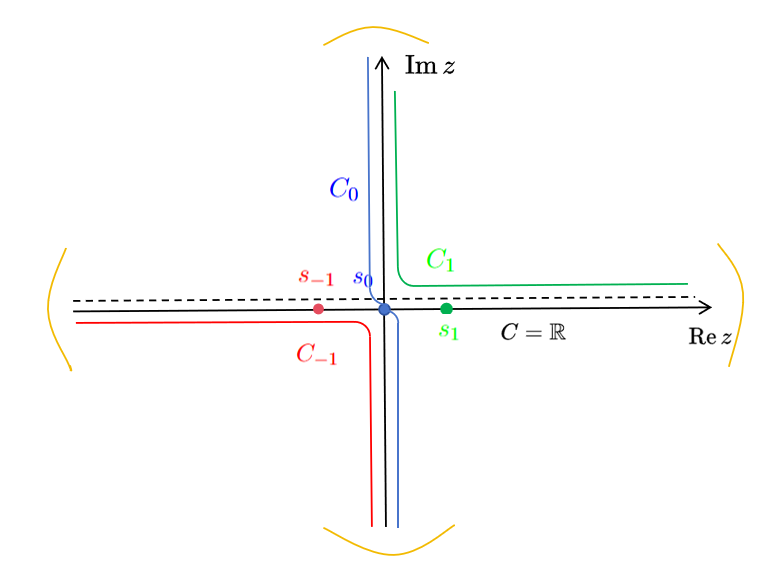}
    \caption{The saddle points of Eq.~\eqref{examplepotential} and a choice of the steepest descent contours in the complexified $z$-plane. Here, the yellow parentheses represent the convergent regions of the integral, while the different colored paths correspond to the integral contours passing through different saddle points.}
    \label{fig:steepest_descent_contour}
\end{figure}

To illustrate the process clearly, consider a specific example. Suppose the action takes the form
\begin{equation}
    S(z) = -\frac{z^2}{2} + \frac{z^4}{4}\, ,\label{examplepotential}
\end{equation}
and we want to integrate it as in Eq.~\eqref{qmpath} along the real line. This action has three (complex) saddle points at $z=-1, 0, 1$, labeled as $s_{-1},s_0,s_1$. The method then instructs us to find the Lefschetz thimbles associated with each saddle point. Fig.~\ref{fig:steepest_descent_contour} shows the saddle points in the complex plane and a choice of the steepest descent contour for each saddle point, in which the parentheses represent the convergent regions of the path integral. Simultaneously, the dashed line in the figure represents the original integral contour $C = \mathbb{R}$. From the figure, we can observe that $C_{-1}$ is the path that originates from the left extremity of the real line, passes through the saddle point $s_{-1}$, reaches the origin, and then proceeds downward along the imaginary axis toward ${\rm Im}z \rightarrow -\infty$, satisfying the requirement of the steepest descent contour. The other two paths can be interpreted in the same manner. We hereby briefly explain why the combination of the three contours is homologous to the original one: since the entire complex plane is effectively spherical, the endpoint of the contour $C_{-1}$ at ${\rm Im}z \rightarrow - \infty$ is connected to the starting point of $C_0$ at ${\rm Im}z \rightarrow \infty$. A similar connection also occurs between the paths $C_{0}$ and $C_1$. Thus, the sum of these three components, $ C_{-1}+ C_0 +C_1$, forms a trajectory that originates from ${\rm Re}z \rightarrow - \infty$, traverses through the three saddle points, and ultimately terminates at ${\rm Re}z \rightarrow \infty$. It is evident that this path can be obtained through a continuous deformation of the original path $C = \mathbb{R}$, implying that they belong to the same equivalent class within the homology group of the manifold. Therefore, as proposed in the Picard-Lecschetz theory, the integral along the real line can be approximated by the combination of the three Lefschetz thimbles, i.e., $\int_C \approx \int_{C_{-1}}+ \int_{C_0} + \int_{C_1}$, and the path integral can be formulated as
\begin{align}
\mathcal{Z} = \mathcal{J}_{-1} + \mathcal{J}_{0} +\mathcal{J}_1 \sim \mathcal{I}_{-1} + \mathcal{I}_{0} + \mathcal{I}_1 \,. 
\label{zformula}
\end{align}
where the integral of each thimble can be calculated through Eq.~\eqref{approxma}, the answer is
\begin{equation}
    \mathcal{I}_{-1} = \sqrt{\pi} \, {\rm exp}\left(\frac{1}{4}\right)\,, \,\,\, \mathcal{I}_0 = \sqrt{-2\pi}\,, \,\,\, \mathcal{I}_{-1} = \sqrt{\pi}\, {\rm exp}\left(\frac{1}{4}\right) \,. \label{real-answer}
\end{equation}
A noteworthy point is that the sum of these three components does not appear to be real, as the contribution of $\mathcal{I}_0$ is imaginary. However, this arises because we have used $\mathcal{I}_i$ to approximate the exact integral $\mathcal{J}_i$, rather than the result of applying the saddle-point expansion. If we rigorously calculate the integral through $\mathcal{J}_i$, although the process for $\mathcal{J}_{-1}$ will generate an imaginary part, the contribution from the other component $\mathcal{J}_1$ along the imaginary axis cancels it exactly, guaranteeing that the full path integral remains real. We provide a detailed discussion of this cancellation process in Appendix~\ref{sec:cancellation}, where we also calculate the sum of different integrals exactly along all Lefschetz thimbles, which is
% ------------------------------------------------------------------------------------------------------------
\begin{equation}
     \mathcal{J}_{-1} + \mathcal{J}_{0} +\mathcal{J}_1 \approx \exp\left(\frac{1}{4}\right) \left(\sqrt{\pi} + 0.75\right)+\sqrt{2\pi}\,,
\end{equation}
Subsequently, whenever we apply the saddle-point expansion method to compute integrals, we will always use the exact integral $\mathcal{J}_{i}$ rather than $\mathcal{I}_i$. In Sec.~\ref{sec:example}, we introduce a general approach that employs the gradient flow equation to find the steepest descent contours.
%===========================================================================================================
\subsection{Quantum field theory case}

The saddle-point expansion method for computing the Euclidean path integral used in quantum field theory can be readily derived by drawing a simple analogy to quantum mechanics, as long as we generalize the coordinate space to the field space. We are still concerned about the case that the potential contains both true vacuum $\phi_-$ and false vacuum $\phi_+$. Thus the path integral we need to compute is
\begin{equation}
    \mathcal{Z} \equiv \left\langle \phi_+ |  e^{- H \mathcal{T}} | \phi_+ \right\rangle = \int \mathcal{D}\phi \,  e^{- S_{E}[\phi]},
    \label{qftpath}
\end{equation}
with the boundary condition $\phi(-\mathcal{T}/2) = \phi(\mathcal{T}/2) = \phi_+$. In which the action now is 
\begin{equation}
    S_{E}[\phi] = \int_{-\mathcal{T}/2}^{\mathcal{T}/2} {\rm d}\tau \int{\rm d}^3 x \left[ \frac{1}{2}\left( \frac{\partial \phi}{\partial \tau} \right)^2 + \frac{1}{2}(\nabla\phi)^2 + V(\phi) \right]\,.
\end{equation}
As before, we can still expand the field $\phi$ around the saddle point of the action, $\phi =\bar{\phi} + \varphi$. Now the classical equation of motion is
\begin{equation}
    \frac{\partial^2\bar\phi}{\partial \tau^2}+{\nabla}^2\bar\phi-V'(\bar\phi)=0\,.
\end{equation}
In the case of quantum field theory, there are still three types of saddle points of $S_{E}[\phi]$: false vacuum static (FV), bounce (B) and shot (S). Substitute the field expansion into the path integral, we can get
\begin{align}
 \mathcal{Z} = \int \mathcal{D}\phi \,  e^{-S_{E}[\phi]} &\approx \sum_{n=FV,B,S} \int_{C_{n}} \mathcal{D} \varphi_n \,  e^{-S_{E}[\bar\phi_n] -\frac{1}{2}S''_{E}[\bar\phi_n]\varphi^{2}_{n}} \,, 
\end{align}

where in the second step we have deformed the integral path into the steepest descent contours passing through the saddle points (labeled by $C_n$). This is because, in the case of quantum field theory, there is still a negative mode corresponding to the bounce solution (The exact bounce solution for some given potential can be found in Refs.~\cite{Lee:1985uv, FerrazdeCamargo:1982sk, Dutta:2011rc, Aravind:2014pva}). Since quantum field theory is just quantum mechanics, with the Hilbert space replaced by the infinite-dimensional Fock space, the application of the steepest descent contour method is not different in both cases. Therefore, we can adopt the same argument as in Sec.~\ref{qmcase}.  Thus, the answer can be written as
% ------------------------------------------------------------------------------------------------------------
\begin{align}
  \mathcal{Z} = \mathcal{J}_{FV} +\mathcal{J}_B + \mathcal{J}_{S} \sim \mathcal{I}_{FV} + \mathcal{I}_{B} + \mathcal{I}_{S}\,.
\end{align}
Note that in false vacuum decay theory, we should usually compare the contributions of the three components to the total path integral, with the bounce part $\mathcal{I}_{\rm B}$ ultimately being the dominant term. A detailed discussion of this process can be found in \cite{Andreassen:2016cvx, Croon:2023zay}. Since the application of the saddle-point expansion in this work does not involve multiple saddle points, we can safely omit this part without compromising rigor.
% -----------------------------------------------------------------------------------------------------------------
\section{Calculating the EFT likelihood}
\label{sec:saddleofLSS}

We now begin to explore the possibility of implementing the saddle-point expansion method to the path integral approach to the LSS. Our starting point is the special relationship between the EFT likelihoods $\mathcal{P}[\delta]$, $\mathcal{P}[\delta_g,\delta]$ and the partition functions $Z[J]$, $Z[J_g,J]$. Building on the previous work \cite{Cabass:2019lqx}, we would like to extend the discussion to second-order terms near the saddle points, with a focus on the treatment of the negative modes. We aim to provide a general approach for calculating the EFT likelihood, taking into account various effects, such as the galaxy stochasticity \cite{Carroll:2013oxa, Rubira:2024tea} and the primordial non-Gaussianity \cite{Nikolis:2024kbx}. 

\subsection{Set up the path integral formula}
\label{sec:LSS-integral-formula}
We first consider the path integral approach to the matter and joint likelihoods. After inheriting the generalization of the predecessors \cite{Carroll:2013oxa,Cabass:2019lqx,Rubira:2023vzw,Rubira:2024tea,Nikolis:2024kbx} (see also some functional methods in Refs.~\cite{Blas:2015qsi, Blas:2016sfa, McDonald:2017ths}), we write the partition function for the correlation functions of the matter field as
% ----------------------------------------------------------------------------------------------------------
\begin{equation}
    Z[J] = \int \mathcal{D}\delta_{\rm in} \, {\rm exp} \left\{ \int_{\boldsymbol{k}} \left(\frac{1}{2}P_{\epsilon_{m}}(k)J(\boldsymbol{k})J(-\boldsymbol{k}) +J(\boldsymbol{k})\delta_{\rm fwd}[\delta_{\rm in}](-\boldsymbol{k})+\cdots\right) \right\}\mathcal{P}[\delta_{\rm in}]\,, \label{matter-partition-function}
\end{equation}
where $\delta_{\rm in}$ is the initial density field and $\mathcal{P}[\delta_{\rm in}]$ is the corresponding initial matter density likelihood. $P_{\epsilon_m}$ denotes the power spectrum of the noise in the matter density field, generated by the loop diagrams \cite{Cabass:2019lqx}. The $\cdots$ in Eq.~\eqref{matter-partition-function} represents higher-order contributions, specifically the terms proportional to the $m$-th ($m>2$) power of $J$. If we assume that the original density probability distribution is Gaussian, then we have
\begin{equation}
    \mathcal{P}[{\delta_{\rm in}}] = {\rm exp}\left\{-\frac{1}{2}\int_{\boldsymbol{k}}\frac{\delta_{\rm in}(\boldsymbol{k})\delta_{\rm in}(-\boldsymbol{k})}{P_{\rm in}(k)}\right\}\,,
\end{equation}
where $P_{\rm in}(k)$ is the initial power spectrum. We will always follow this assumption in the subsequent parts of this paper. Analogously, we can construct the partition function for the joint case in the same way, except for two differences: First, there are now two currents $J_g$, $J$ associated with the galaxy field $\delta_g$ and the matter field $\delta$, respectively; Second, the previous noise term now needs to be generalized into three components, corresponding to (a) the stochasticity for galaxies $P_{\epsilon_g} \sim k^0$, (b) the cross stochasticity between galaxies and matter $P_{\epsilon_g \epsilon_m} \sim k^2$, and (c) the matter stochasticity $P_{\epsilon_m} \sim k^4$ (where $\sim$ represents their leading-order contributions). Thus, the partition function for the joint case is
\begin{align}
    Z[J_g,J] &= \int \mathcal{D} \delta_{\rm in} \, {\rm exp}\left\{\int_{\boldsymbol{k}}\left[\frac{1}{2}P_{\epsilon_g}(k)J_g(\boldsymbol{k})J_g(-\boldsymbol{k}) + P_{\epsilon_g\epsilon_m}(k)J_g(\boldsymbol{k})J(-\boldsymbol{k}) \right]\right\} \nonumber \\ 
     & \times {\rm exp}\left\{\int_{\boldsymbol{k}}\left[\frac{1}{2}P_{\epsilon_m}(k)J(\boldsymbol{k})J(-\boldsymbol{k}) +J_g(\boldsymbol{k})\delta_{\rm g,fwd}[\delta_{\rm in}](-\boldsymbol{k}) +J(\boldsymbol{k})\delta_{\rm fwd}[\delta_{\rm in}](-\boldsymbol{k}) + \cdots   
     \right]\right\} \mathcal{P}[\delta_{\rm in}]\,.
\end{align}
Also at this point, $\cdots$ represents higher-order terms. A noteworthy point is that all the stochastic terms stem from the coarse-grained matter field, as these terms are required to counteract the UV dependence of the loop diagrams. Therefore, we need to regularize the integral over $\mathcal{D}\delta_{\rm in}$ by introducing a hard cutoff in the Fourier space. The relevant content has been illustrated in detail in the previous literatures \cite{Peskin:1995ev,Cabass:2019lqx}, we will skip this discussion and follow the convention of \cite{Cabass:2019lqx}.

A key advantage of the path-integral formulation is that it allows us to exploit all available information. In \cite{Cabass:2019lqx}, the authors calculated the conditional likelihood using this approach and demonstrated that, under the same assumptions --- such as the Gaussian initial condition and two nonvanishing stochasticities $P_{\epsilon_{g}}$ and $P_{\epsilon_{g}\epsilon_{m}}$ --- the path-integral approach not only reproduces the results of \cite{Schmidt:2018bkr}, but also recovers certain terms that were omitted in that work.

The origin of these additional terms can be attributed to a drawback of the standard perturbation theory (SPT): when computing the $n$-point correlation functions, correlators of order higher than $n$ are involved. This can be seen, for example, in the equations referred to as the BBGKY hierarchy in \cite{Peebles:1980yev}. At the same time, from the continuity equation for the probability density in \cite{Blas:2015qsi} (where this approach is also called the time-sliced perturbation theory) and the renormalization group equations in Refs.~\cite{Rubira:2023vzw, Rubira:2024tea, Nikolis:2024kbx}, it is evident that in this path-integral formulation, the $n$-th order evolution equation will not contain terms higher than $n$. Therefore, we conclude that these extra terms are indeed missing in \cite{Schmidt:2018bkr}, and the functional method provides more precise results.

In this approach, the likelihood is simply the functional Fourier transform of the partition function. If we define the functional Dirac function with a normalization factor $\mathcal{N}_{\delta^{(\infty)}}$,
\begin{equation}
    \delta_{D}^{(\infty)}(\varphi-\chi) = \mathcal{N}_{\delta^{(\infty)}} \, \int \mathcal{D}X \,{\rm exp}\left\{i \int_{\boldsymbol{k}}X(\boldsymbol{k})(\varphi(-\boldsymbol{k})-\chi(-\boldsymbol{k}))\right\}\, . \label{functional-Dirac}
\end{equation}
We can then derive the functional formula for the matter likelihood using Eq.~\eqref{matter-partition-function}
\begin{equation}
    \mathcal{P}[\delta] =\mathcal{N}_{\delta^{(\infty)}}\, \int \mathcal{D} X \, {\rm exp} \left\{i\int_{\boldsymbol{k}}X(\boldsymbol{k})\delta(-\boldsymbol{k})\right\} Z[-iX]\,, \label{matter-likelihood}
\end{equation}
where we have written the current corresponding to $\delta$ as $X = iJ$. Similarly, we can obtain the expression for the joint likelihood
\begin{equation}
    \mathcal{P}[\delta_g,\delta] = \mathcal{N}_{\delta^{(\infty)}}^{2} \int\mathcal{D}X_g \mathcal{D}X \, {\rm exp}\left\{i\int_{\boldsymbol{k}} \left[X_g(\boldsymbol{k}) \delta_g(-\boldsymbol{k})+X(\boldsymbol{k})\delta(-\boldsymbol{k})\right]\right\} Z[-iX_g,-iX], 
    \label{joint-likelihood}
\end{equation}
where $X_g = iJ_g$ represents the current for the galaxy. The main goal of this work is to calculate these two likelihoods Eqs.~\eqref{matter-likelihood} and \eqref{joint-likelihood} through the saddle-point expansion. In the next subsection, we will explore the role of saddle-point expansion in this path-integral formulation.

% ---------------------------------------------------------------------------------------------------------------
\subsection{The saddle-point expansion}
To apply the saddle-point expansion, we need an expression similar to the forms of Eqs.~\eqref{qmpath} and ~\eqref{qftpath}. This can be easily achieved by rewriting the expressions for the two likelihoods. We first consider the matter likelihood, if we package all the integral parameters into a field $\boldsymbol{\phi}$, satisfying
\begin{equation}
    \boldsymbol{\phi} = \begin{pmatrix}
        X \\ \delta_{\rm in}
    \end{pmatrix},
\end{equation}
then the likelihood can be expressed as
\begin{equation}
    \mathcal{P}[\delta] = \int \mathcal{D} \boldsymbol{\phi} \,  e^{-S[\boldsymbol{\phi}]}, \label{mattercase}
\end{equation}
where we have included all the terms in the exponent as the action. Note that in this case, we do not impose similar boundary conditions as in false vacuum decay theory. Despite being wordy, we still emphasize here that the boundary conditions are crucial in false vacuum decay theory, because the path integral is the transition matrix between two states. We need $\phi = \phi_+$ at both $\mathcal{T}/2$ and $-\mathcal{T}/2$\footnote{In fact, the Minkowski time ${T}$ should be sufficiently large but not tend to infinity in false vacuum decay theory. We need large ${T}$ to ensure the decay process has fully taken place, and ${T}\rightarrow \infty$ is also forbidden because if so we will instead capture the contribution from the lowest energy state which is the true vacuum, as illustrated in \cite{Ai:2019fri}. This argument is also consistent with the numerical results in quantum mechanics, which shows the decay rates have no exponential form for either large ${T}$ and small ${T}$ \cite{Andreassen:2016cvx}.} to ensure the transition is from the false vacuum state to itself. However in this case there is neither time $T$ in the expression for Eq.~\eqref{mattercase}, nor the answer corresponds to some transition matrix, so we can ignore this.

In this way, the most general formula of the action can be expressed as
\begin{equation}
    S[\boldsymbol{\phi}] = \int_{\boldsymbol{k}} \boldsymbol{\phi}^{a}(\boldsymbol{k}) \mathcal{J}^{a}(-\boldsymbol{k}) + \frac{1}{2}\int_{\boldsymbol{k},\boldsymbol{k}'} \mathcal{M}^{ab}(\boldsymbol{k},\boldsymbol{k}') \boldsymbol{\phi}^a(\boldsymbol{k}) \boldsymbol{\phi}^b(\boldsymbol{k}') + \frac{1}{3!}\int_{\boldsymbol{k},\boldsymbol{k}',\boldsymbol{k}''}\mathcal{M}^{abc} \boldsymbol{\phi}^a\boldsymbol{\phi}
^b\boldsymbol{\phi}^c +\cdots , \label{matter-action}
\end{equation}
where we can read off the expression for $\mathcal{J}$ through Eq.~\eqref{matter-likelihood}, $\mathcal{J} = (i\delta,0)$, and the remaining matrix expressions can be extracted in the same manner. There are three things that should be emphasized about the above equation: First, in contrast to \cite{Cabass:2019lqx}, the linear term of $\boldsymbol{\phi}$ has been incorporated into the action. Note that when performing the saddle-point expansion, the saddle point must be computed by considering all terms in the exponent, therefore under the former convention the saddle point of the partition function corresponding to the connected diagrams is required. Second, when calculating the decay rates, the first step is to transform to Euclidean space. The equation of motion in Euclidean space admits an additional bounce solution compared to the Minkowski spacetime. This solution becomes the dominant contribution to the path integral and does not appear in the normalization constant \cite{Callan:1977pt}. However, in the present case, we compute the likelihood directly in the three-dimensional Euclidean space, so there is no need to consider the issue of the extra instanton solution. In the following calculations, we will also omit the normalization factor $\mathcal{N}_{\delta^{(\infty)}}$ in Eq.~\eqref{functional-Dirac} which is independent of the fields, and recover it at the end. Another point worth noting is that, as seen from Eq.~\eqref{matter-likelihood} and Eq.~\eqref{joint-likelihood}, the expression for the actions may not be real, since the imaginary unit $i$ appears in the functional Fourier transformation. Therefore, one may think that the results of the likelihoods may not be real. However, we reiterate here that the likelihoods are real not only because they need to correspond to the probability, but also, as we will demonstrate in the following example in Sec.~\ref{sec:example}, if we use the gradient flow equation to solve for the steepest descent contours, all eigenvalues will also lead to a real conditional likelihood. 

Now that we have the consistent formula, we can compute this ``path integral" as we did in Sec.~\ref{sec:qmandqft}. We decompose the field $\boldsymbol{\phi}$ into two parts $\boldsymbol{\phi} = \bar{\boldsymbol{\phi}} + \boldsymbol{\varphi}$, in which $\bar{\boldsymbol{\phi}}$ is the saddle point of the action, satisfying
\begin{equation}
    \frac{\delta S[\boldsymbol{\phi}]}{\delta\boldsymbol{\phi}}\bigg|_{\boldsymbol{\phi} = \bar{\boldsymbol{\phi}}} = 0\,,
\end{equation}
and $\boldsymbol{\varphi}$ is the ``quantum fluctuation". Then the matter likelihood can be approximated as
\begin{equation}
    \mathcal{P}[\delta] = \int \mathcal{D} \boldsymbol{\phi} \, e^{-S[\boldsymbol{\phi}]}
\approx \int \mathcal{D} \boldsymbol{\phi} \, e^{-S[\bar{\boldsymbol{\phi}}] - \frac{1}{2}S''[\bar{\boldsymbol{\phi}}]\boldsymbol{\varphi}^2 +\cdots} \sim  e^{-S[\bar{\boldsymbol{\phi}}]}({\rm det}S''[\bar{\boldsymbol{\phi}}])^{-1/2}\, . 
\label{matter-likelihood1}
\end{equation}
If there are negative modes in the computation process, i.e. there are negative eigenvalues of $S''[\bar{\boldsymbol{\phi}}]$, then our calculation will also be ill-defined. At this point, we need to use the same argument as in quantum mechanics and quantum field theory, approximating the integral contour by the combination of the steepest descent contours connect different saddle points and terminate at the convergent regions in the $\boldsymbol{\phi}$ space. Thus we have
\begin{equation}
    \mathcal{P}[\delta] \approx  e^{-S[\bar{\boldsymbol{\phi}}]} \int_C \mathcal{D}\boldsymbol{\varphi}\, e^{-\frac{1}{2}S''[\bar{\boldsymbol{\phi}}]\boldsymbol{\varphi}^2} \, .
\end{equation}
where $C$ is one of the choices of the combination of the steepest descent contours that is homologous to the original integral contour we have chosen. It can be seen that the situation here is analogous to that in quantum field theory when calculating the decay rate with multiple fields \cite{Garbrecht:2015yza}. So after proceeding with this, we can apply the same discussion as in quantum field theory.

Next, we consider the calculation of joint likelihood. It is straightforward that the previous discussions can be easily extended to the case of joint likelihood. We only need to draw an analogy with the matter situation and redefine a ``galaxy field" $\boldsymbol{\phi}_g$, 
\begin{equation}
    \boldsymbol{\phi}_g = \begin{pmatrix}
       X_g \\ X \\ \delta_{\rm in}
    \end{pmatrix},
\end{equation}
and the joint likelihood can be expressed as
\begin{equation}
      \mathcal{P}[\delta,\delta_g] = \int \mathcal{D} \boldsymbol{\phi}_g \,  e^{-S_g[\boldsymbol{\phi}_g]} \, ,
\end{equation}
We can similarly read off the action from Eq.~\eqref{joint-likelihood} 
\begin{equation}
     S_g[\boldsymbol{\phi}_g] = \int_{\boldsymbol{k}} \boldsymbol{\phi}_{g}^{a}(\boldsymbol{k}) \mathcal{J}_{g}^{a}(-\boldsymbol{k}) + \frac{1}{2}\int_{\boldsymbol{k},\boldsymbol{k}'} \mathcal{M}_{g}^{ab}(\boldsymbol{k},\boldsymbol{k}') \boldsymbol{\phi}_{g}^{a}(\boldsymbol{k}) \boldsymbol{\phi}^{b}_{g}(\boldsymbol{k}') + \frac{1}{3!}\int_{\boldsymbol{k},\boldsymbol{k}',\boldsymbol{k}''}\mathcal{M}^{abc}_{g} \boldsymbol{\phi}_{g}^{a}\boldsymbol{\phi}_{g}
^{b}\boldsymbol{\phi}_{g}^{c} +\cdots . \label{joint-action1}
\end{equation}
Then the saddle-point expansion implies
\begin{equation}
        \mathcal{P}[\delta_g,\delta] = \int \mathcal{D} \boldsymbol{\phi}_g \, {\rm}e^{-S_g[\boldsymbol{\phi}_g]}
\approx \int_C \mathcal{D} \boldsymbol{\varphi}_g \, e^{-S_g[\bar{\boldsymbol{\phi}}_g] - \frac{1}{2}S_g''[\bar{\boldsymbol{\phi}}_g]\boldsymbol{\varphi_g}^2 +\cdots} \sim  e^{-S_g[\bar{\boldsymbol{\phi}}_g]}({\rm det}S_g''[\bar{\boldsymbol{\phi}}_g])^{-1/2}, 
\end{equation}
where we have decomposed the field $\boldsymbol{\phi}_g$ into $\boldsymbol{\phi}_g =\bar{\boldsymbol{\phi}}_g +\boldsymbol{\varphi}_g$ and implemented the steepest descent contour $C$ as before. It can be observed that it is always necessary to discuss the eigenvalues of the second derivative of the action. Fortunately, in contrast to the false vacuum decay theory, when solving for the likelihood the expression for $S''[\bar{\boldsymbol{\phi}}]$ can often be written as a determinate matrix, which significantly reduces the computational workload. 

Another naive problem is whether it is justified to use the saddle-point expansion in the path-integral approach to the LSS. In quantum field theory, if we recover $\hbar$, then the field decomposition should be $\phi =\bar{\phi} + \hbar^{1/2}\varphi$, such that the contribution near the saddle point is of order $\mathcal{O}(\hbar)$ (i.e. proportional to $\hbar\,\varphi^{2}$), and all higher-order terms are suppressed by powers of $\hbar$. However now the path-integral approach to the LSS is not a quantum theory, so there is no $\hbar$ in the expression for the decomposition of $\boldsymbol{\phi}$ or $\boldsymbol{\phi}_g$, nor is there a similar perturbative parameter in this theory which can be used to draw an analogy. 

We hereby offer a justification for using saddle points to approximate the full path integral. The rationale can be traced to the method of stationary phase (also known as the method of steepest descent) in quantum field theory (see, for example, Sec.~14.2 of \cite{Schwartz:2014sze}). Suppose we consider a path integral of the following form
% ---------------------------------------------------------------------------------------------------------------
\begin{equation}
    \int\mathcal{D} \Phi (\boldsymbol{x},t) \, e^{-\frac{1}{\hbar}S[\Phi]}.
\end{equation}
In the classical limit $\hbar \rightarrow 0$, this integral is dominated by the values of $\Phi$ for which $S[\Phi]$ has an extremum, i.e. $\delta S[\Phi] = 0$, because these values correspond to the regions where the action oscillates at least. This is why we can decompose the field as $\Phi = \bar{\Phi} + \mathcal{O}(\hbar)\,\phi$, where $\bar{\Phi}$ is the stationary point of the action. We expect the same phenomenon to occur in our case. This can be achieved by performing a rescaling analogous to the one in \cite{Cabass:2019lqx}, taking $\boldsymbol{k} = b \boldsymbol{k}'$ and setting the ``external currents" $\delta$ and $\delta_g$ to zero. Therefore, in the limit $b \rightarrow 0^+$, the rescaling of different operators will reflect their relevance (or irrelevance) to the large-scale universe. In this context, we employ $b$\footnote{Note that here $b$ is not a perturbative parameter but a rescaling parameter.} as a parameter analogous to $\hbar$ in quantum mechanics. Under this rescaling, the integration measure ${\rm d}^3 k$ changes to $b^3{\rm d}^3k'$, and the power spectrum $P_{\rm in}$ scales as $P_{\rm in}(k) = b^{n_{\delta}}P_{\rm in}(k')$, where $n_{\delta}$ is the order of momentum in the initial power spectrum. Then, the terms in the action that are quadratic in $X_g$ and $\delta_{\rm in}$ remain invariant if we redefine
% ------------------------------------------------------------------------------------------------------------
\begin{align}
    X_g(b\boldsymbol{k}') &= b^{-\frac{3}{2}}X_{g}'(\boldsymbol{k'}) \,,\\
    \delta_{\rm in} (b\boldsymbol{k}') &= b^{\frac{n_\delta-3}{2}} \delta_{\rm in} (\boldsymbol{k}')\,.
\end{align}
For $n_{\delta} = -1.7$, it can be observed that the rescaling rate of $\delta_{\rm in}$ is faster than $X_g$. In this case, if we only consider the lowest-order nonvanishing terms in the action Eq.~\eqref{joint-action1}, then the rescaling behavior of the action is
% -------------------------------------------------------------------------------------------------------------
\begin{equation}
    S_g[\boldsymbol{\phi}_g] \rightarrow b^{-3} S_g[\boldsymbol{\phi}_g]\,,
\end{equation}
and the path-integral should to be rescaled as
% ---------------------------------------------------------------------------------------------
\begin{equation}
\int \mathcal{D} \boldsymbol{\phi}_g \, {\rm}e^{-S_g[\boldsymbol{\phi}_g]} \rightarrow \int \mathcal{D} \boldsymbol{\phi}_g \, {\rm}e^{-\frac{1}{b^3}S_g[\boldsymbol{\phi}_g]}\,.
\end{equation}
We can now draw an analogy with the situation in quantum field theory: at large scales ($b\rightarrow 0^+$), the saddle-point contribution to the action will dominate the path integral, and other terms are suppressed. Thus, we have demonstrated that it is convincing to approximate the path integral using the saddle points. Then, the field expansion should be
% -------------------------------------------------------------------------------------------------------
\begin{equation}
    \boldsymbol{\phi}_g =\bar{\boldsymbol{\phi}}_g +\mathcal{O}(b)\,\boldsymbol{\varphi}_g\,.
\end{equation}
Furthermore, the contributions near the saddle points can then be appropriately interpreted as the second-order corrections to the likelihoods. We acknowledge that the functional approach based on the path integral provides more accurate results, and our method will further advance this framework. In the subsequent section, we will illustrate that the application of the saddle-point expansion introduces an additional factor in the likelihood result than before, which aligns with our expectations.

%===============================================================================================================
\section{Example: the EFT likelihood with a Gaussian distribution}
\label{sec:example}
Upon establishing the appropriate theoretical framework, it can be applied to specific models. In this section, we focus on the model presented in \cite{Cabass:2019lqx}, which involves the computation of the conditional likelihood with a Gaussian distribution. We then demonstrate the application of the steepest descent contour method through concrete examples. 

As outlined in Sec.~\ref{sec:LSS-integral-formula}, there are three stochastic terms for matter and galaxy, $P_{\epsilon_g}$, $P_{\epsilon_g\epsilon_m}$, $P_{\epsilon_m}$. In the case where only $P_{\epsilon_g}$ exists, the likelihoods can be evaluated exactly, yielding rigorous results. This example has been previously investigated in \cite{Schmidt:2018bkr, Cabass:2019lqx}. Hence, we consider the more intricate case where both $P_{\epsilon_g}$ and $P_{\epsilon_g\epsilon_m}$ are non-vanishing, and demonstrate that the saddle-point expansion yields identical conditional likelihood.

We begin by examining the structure of the actions. After incorporating the aforementioned approximations, the actions now contain only terms up to the second order in $J$ and $J_g$, i.e. up to $\boldsymbol{\phi}^2$ and $\boldsymbol{\phi}_{g}^{2}$ order. The expression for the ``matter action" is
\begin{equation}
    S[\boldsymbol{\phi}] = \int_{\boldsymbol{k}} \boldsymbol{\phi}^{a}(\boldsymbol{k}) \mathcal{J}^{a}(-\boldsymbol{k}) + \frac{1}{2}\int_{\boldsymbol{k},\boldsymbol{k}'} \mathcal{M}^{ab}(\boldsymbol{k},\boldsymbol{k}') \boldsymbol{\phi}^a(\boldsymbol{k})\boldsymbol{\phi}^b(\boldsymbol{k}')\, , \label{matter-action}
\end{equation}
and for the ``joint action"
\begin{equation}
      S_g[\boldsymbol{\phi}_g] = \int_{\boldsymbol{k}} \boldsymbol{\phi}^{a}_{g}(\boldsymbol{k}) \mathcal{J}^{a}_{g}(-\boldsymbol{k}) + \frac{1}{2}\int_{\boldsymbol{k},\boldsymbol{k}'} \mathcal{M}^{ab}_{g}(\boldsymbol{k},\boldsymbol{k}') \boldsymbol{\phi}^{a}_{g}(\boldsymbol{k})\boldsymbol{\phi}^{b}_{g}(\boldsymbol{k}')\, . \label{joint-action}
\end{equation}
We can read off the formula of $\mathcal{M}^{ab}$ and $\mathcal{M}_{g}^{ab}$ from the expression for the likelihoods, i.e. Eq.~\eqref{matter-likelihood} and Eq.~\eqref{joint-likelihood}
\begin{equation}
    \mathcal{M}(\boldsymbol{k},\boldsymbol{k}') = (2\pi)^3 \delta_{D}^{(3)}(\boldsymbol{k}+\boldsymbol{k}') \begin{pmatrix}
      0 & i K_1(k)D_1 \\
i K_1(k)D_1 & P^{-1}_{\rm in}(k)
    \end{pmatrix}\,, \label{m-form}
\end{equation}
and
\begin{equation}
      {\cal M}_g(\boldsymbol{k},\boldsymbol{k}') = (2\pi)^3\delta^{(3)}_{D}(\boldsymbol{k}+\boldsymbol{k}')
\begin{pmatrix}
P_{\epsilon_g}(k) & P_{\epsilon_g\epsilon_m}(k) & i K_{g,1}(k) D_1 \\
P_{\epsilon_g\epsilon_m}(k) & 0 & i K_1(k)D_1 \\
i K_{g,1}(k) D_1 & i K_1(k)D_1 & P^{-1}_{\rm in}(k)
\end{pmatrix}\,\,.\label{mg-form}
\end{equation}
At zeroth order, we can take $K_1 = 1$ and $K_{g,1} = b_1$. We briefly review how to calculate the contribution of the saddle points, with the detailed results provided in Appendix~\ref{sec:saddle-point-contribution}, and here we focus on the issues related to the negative modes.

We first derive the equations of motion for the matter and galaxy fields. By taking the functional derivatives with respect to the fields  $\boldsymbol{\phi}$ and $\boldsymbol{\phi_g}$, we obtain
\begin{align}
    \frac{\delta S}{\delta \boldsymbol{\phi}}\bigg|_{\boldsymbol{\phi} = \bar{\boldsymbol{\phi}}} = \mathcal{J}^{a} + \mathcal{M}^{ab}\bar{\boldsymbol{\phi}} =0 \,,
    \\
     \frac{\delta S_g}{\delta \boldsymbol{\phi}_g}\bigg|_{\boldsymbol{\phi}_g = \bar{\boldsymbol{\phi}}_g} = \mathcal{J}^{a}_{g} + \mathcal{M}^{ab}_{g}\bar{\boldsymbol{\phi}}_{g} =0 \,.
\end{align}
Note that due to the presence of the Dirac delta function in the expressions for $\mathcal{M}$ and $\mathcal{M}_g$, we need not consider the case when $\boldsymbol{k} \neq \boldsymbol{k}'$. The expressions for the solutions of these two equations and $-S[\bar{\boldsymbol{\phi}}]$, $-S_{g}[\bar{\boldsymbol{\phi}}_g]$ can be found in Appendix~\ref{sec:saddle-point-contribution}. We now consider the additional terms arising from the saddle-point expansion. For both the matter and galaxy case, $S''[\bar{\boldsymbol{\phi}}]$ and $S''_g[\bar{\boldsymbol{\phi}}_g]$ are simply the matrices corresponding to the second-order terms in the fields, $\mathcal{M}^{ab}$ and $\mathcal{M}^{ab}_{g}$. For the matter likelihood, the eigenvalues of $S''[\bar{\boldsymbol{\phi}}]$ can be simply calculated, 
\begin{equation}
    m_1 = \frac{1}{2}(P_{\rm in}^{-1} - \sqrt{-4 D_{1}^{2} +P_{\rm in}^{-2}}) \, , \,\,\,\,\,\, m_2 =  \frac{1}{2}(P_{\rm in}^{-1} + \sqrt{-4 D_{1}^{2} +P_{\rm in}^{-2}})\,.
\end{equation}
It is evident that provided the two eigenvalues are real (the case of the complex eigenvalues will be addressed below), they must be positive because of the positivity of $P_{\rm in}$ and $D_{1}$, as shown by the equation $P_{\rm in}^{-1} - \sqrt{-4 D_{1}^{2} +P_{\rm in}^{-2}} = \sqrt{P_{\rm in}^{-2}}-  \sqrt{-4 D_{1}^{2} +P_{\rm in}^{-2}} $. Therefore, for the matter likelihood, there are no issues related to the negative modes, and we can directly proceed with Eq.~\eqref{matter-likelihood1} to obtain
% ------------------------------------------------------------------------------------------------------------
\begin{equation}
    \mathcal{P}[\delta] \approx   e^{-S[\bar{\boldsymbol{\phi}}]}({\rm det}S''[\bar{\boldsymbol{\phi}}])^{-1/2} =\sqrt{\pi}\,(\prod_i m_i)^{-1/2} \, {\rm e}^{-S[\bar{\boldsymbol{\phi}}]}=\frac{\sqrt{\pi}}{D_1}\,  e^{-S[\bar{\boldsymbol{\phi}}]}. 
    \label{matter-answer}
\end{equation}
It can be seen that after incorporating the second-order expansion of the saddle points, the resulting expression is equivalent to the original one with the addition of an extra factor, as anticipated. Analogously, the calculation of joint likelihood should be obtained in the same manner, however, we will encounter certain intractable problems now: Given a general three-dimensional matrix $\mathcal{M}_g$, there are no simple analytical expressions for its eigenvalues. Even when we obtain them through complicated derivations, the solutions are usually not positive, and even not real because $\mathcal{M}_g$ is neither Hermitian nor anti-Hermitian. At this time, there are no shortcuts, and we need to follow the previous discussions regarding the negative modes. We need to complexify the field $\boldsymbol{\phi}_g$ and then seek the approximate contour that is homologous to the original one through the gradient flow equation. At the same time, there is no zero mode corresponding to $S''_g[\bar{\boldsymbol{\phi}}_g]$, so we do not need to be concerned about the issues related to the collective coordinates.

We will present a general method for solving this type of problem, and the following discussion is derived by analogy with \cite{Ai:2019fri}. We start from the complexified $\boldsymbol{\phi}$ space. We introduce the real parameter $u$ to label different paths passing through the saddle points in the complex plane, and impose the boundary condition $\boldsymbol{\phi}_g(\boldsymbol{k},u \rightarrow -\infty) = \bar{\boldsymbol{\phi}}_g$, i.e. under this parametrization, in the limit of $u \rightarrow - \infty$, we recover the saddle points. Since our goal is to identify the steepest descent contours, the holomorphic function we consider should be the negative value of the action $\mathcal{I}[\boldsymbol{\phi}_g] = - S_g[\boldsymbol{\phi}_g]$, and then define the Morse function $f[\boldsymbol{\phi}_g] = {\rm Re}(\mathcal{I}[\boldsymbol{\phi}_g])$. Thus, for the saddle point $\bar{\boldsymbol{\phi}}_g$, the steepest descent path is given by the gradient flow equation \cite{Witten:2010cx}
\begin{equation}
    \frac{\partial \boldsymbol{\phi_g}(\boldsymbol{k},u)}{\partial u} = - \overline{\left( 
    \frac{\delta\mathcal{I}[\boldsymbol{\phi}_g(\boldsymbol{k},u)]}{\delta\boldsymbol{\phi}_g(\boldsymbol{k},u)}\right)}, \ \  \ \ { \frac{\partial \overline{\boldsymbol{\phi_g}(\boldsymbol{k},u)}}{\partial u} }= - \left( 
    \frac{\delta\mathcal{I}[\boldsymbol{\phi}_g(\boldsymbol{k},u)]}{\delta\boldsymbol{\phi}_g(\boldsymbol{k},u)}\right), \label{gradient-flow-equation}
\end{equation}
where $u \in \mathbb{R}$ and the boundary condition is $\boldsymbol{\phi}_{g} (\boldsymbol{k}, u\rightarrow-\infty) = \bar{\boldsymbol{\phi}}_g$.  The overline represents the complex conjugate. We can easily check that
\begin{equation}
\frac{\partial f}{\partial u} = \frac{1}{2}(\frac{\delta \mathcal{I}}{\delta \boldsymbol{\phi}_g}\frac{\partial \boldsymbol{\phi}_g}{\partial u} + 
\frac{\delta\bar{\mathcal{I}}}{\delta\bar{\boldsymbol{\phi}}_g}\frac{\partial\bar{\boldsymbol{\phi}}_g}{\partial u}) = - \bigg|\frac{\partial\boldsymbol{\phi}_g(\boldsymbol{k},u)}{\partial u}\bigg|^2 \leq 0 \,,
\end{equation}
this implies that the real part of $\mathcal{I}[\boldsymbol{\phi}_g]$ is decreasing along the contour as it moves away from the saddle point, fulfilling the condition of the steepest descent contour. Upon plugging the expression for the action into Eq.~\eqref{gradient-flow-equation}, we can get
\begin{equation}
\frac{\partial\boldsymbol{\phi}_g(\boldsymbol{k},u)}{\partial u} = - {\mathcal{J}}^{*}_{g}(\boldsymbol{k}) + {\mathcal{M}}^{*}_{g}{\boldsymbol{\phi}}^{*}_{g} \, .
\end{equation}
This equation corresponds to the steepest descent contour in this context. Next, we decompose the field as $\boldsymbol{\phi}_g = \bar{\boldsymbol{\phi}}_g +\boldsymbol{\varphi}_g$, and apply the complex conjugate of the saddle-point equation: $\mathcal{J}^{*}_{g} +\mathcal{M}^{*}_{g} \bar{\boldsymbol{\phi}}_{g}^{*} = 0 $. Consequently, the equation above can be rewritten as
\begin{equation}
\frac{\partial\boldsymbol{\varphi}_g(\boldsymbol{k},u)}{\partial u} = \mathcal{M}^{*}_{g} \varphi^{*}_{g} \, .
\end{equation}
We can employ the method of separation of variables to reduce this equation by making the following ansatz: $\boldsymbol{\phi}_g(\boldsymbol{k},u) = \sum_n g_{n}(u)\chi_{n}(\boldsymbol{k})$, where $g_{n}(u) \in \mathbb{R}$ and the subscript ``$n$" distinguishes different directions. Substituting this ansatz into the above equation, we obtain the following result
\begin{equation}
    \mathcal{M}_{g}^{*} \, g_{n}(u) \chi_{n}^{*}(\boldsymbol{k}) = \chi_{n}(\boldsymbol{k}) \frac{{\rm d} g_{n}(u)}{{\rm d} u}\, ,
\end{equation}
then we have
\begin{equation}
    \mathcal{M}_{g}^{*} \, \chi_{n}^{*}(\boldsymbol{k})/\chi_{n}(\boldsymbol{k}) = m_{n} =\frac{1}{g_{n}(u)} \frac{{\rm d} g_{n}(u)}{{\rm d} u}\, , 
    \label{full-eigenequation}
\end{equation}
where we have introduced the real eigenvalue $m_{n}$, as all the terms in the RHS  of the equation are real (a rigorous proof of this will be provided below). Thus the eigenequation can be expressed as
\begin{equation}
     \mathcal{M}_{g}^{*} \, \chi_{n}^{*}(\boldsymbol{k}) = m_{n} \,\chi_{n}(\boldsymbol{k}) \,. \label{eigenequation}
\end{equation}
For a general three-dimensional matrix $\mathcal{M}_g^{*}$, obtaining analytical solutions for the eigenvalues is generally challenging. We can use numerical methods to approximate the eigenvalues after obtaining its explicit formula. Another issue occurs during the saddle-point expansion is that what we need to calculate is $S''[\bar{\boldsymbol{\phi}}_g] \boldsymbol{\varphi}_g \sim \mathcal{M}_g \boldsymbol{\varphi}_g$, therefore the relevant equation is not the one presented above, but rather its complex conjugate $\mathcal{M}_{g} \, \chi_{n}(\boldsymbol{k}) = m_{n} \,\chi_{n}^{*}(\boldsymbol{k})$\footnote{In fact, for each eigenvalue $m_n$ there is always an accompanying eigenvalue $-m_n$, corresponding to the eigenstate ${i \,\chi_{n}(\boldsymbol{k})}$. However, as we will demonstrate in the following discussion,  all $m_n$ are real and positive, allowing us to disregard all the $-m_n$.}. Furthermore, in Eq.~\eqref{eigenequation}, the matrix $\mathcal{M}^{*}_g$ is neither Hermitian nor are the eigenfunctions real, which precludes the imposition of the standard normalization condition on the eigenstates. To work out this, we can construct a Hermitian operator by combining Eq.~\eqref{eigenequation} with its complex conjugate
\begin{equation}
        \begin{pmatrix}
         \boldsymbol{0}  & \mathcal{M}^{*}_{g} \\
         \mathcal{M}_{g} & \boldsymbol{0}
        \end{pmatrix}
    \begin{pmatrix}
        \chi_{n}(\boldsymbol{k}) \\ {\chi}_{n}^{*}(\boldsymbol{k})
    \end{pmatrix}
     =  m_{n}\begin{pmatrix}
         \chi_{n}(\boldsymbol{k}) \\ {\chi}_{n}^{*}(\boldsymbol{k})
     \end{pmatrix} \, , \label{hermite-equation}
\end{equation}
thus the operator on the left-hand side is now Hermitian, and the eigenvalues $m_{n}$ should be real. However, this equation remains intractable for an exact solution, and the steps for obtaining an approximate solution are provided in Appendix~\ref{sec:eigenvalues}. In this case, we can also impose the orthonormalization to the eigenstates
\begin{equation}
    \int_{\boldsymbol{k}} {\chi}_{n}^{*}(\boldsymbol{k}) \, \chi_{n}(\boldsymbol{k}) = \delta_{\rm mn}\, .
\end{equation}
On the other hand, for the equation on the RHS of Eq.~\eqref{full-eigenequation}, the solution is expected to take the form $g_{n} (u) \sim a_{n}\,{\rm exp}\,(m_nu)$, where $a_n \in \mathbb{R}$. Using the boundary condition $g(u\rightarrow - \infty) =0$, it follows that $m_n > 0$, i.e. the eigenvalues are all positive and real.

Now we can calculate the joint likelihood
\begin{align}
    \mathcal{P}[\delta_g,\delta] & \approx  \int_{C} \mathcal{D} \boldsymbol{\phi}_g \,  e^{-S_g[\bar{\boldsymbol{\phi}}_g] -\frac{1}{2} \boldsymbol{\varphi}_g S''_g[\bar{\boldsymbol{\phi}}_g]\boldsymbol{\varphi}_g}  = \int_C \mathcal{D} \boldsymbol{\varphi}_g \,  e^{-S_g[\bar{\boldsymbol{\phi}}_g] - \boldsymbol{\varphi}_g M_g\boldsymbol{\varphi}_g}\nonumber\\ 
    & =\int \mathcal{D} \boldsymbol{\varphi}_g \,e^{-S_g[\bar{\boldsymbol{\phi}}_g] - \sum_n m_n|\boldsymbol{\varphi}_g|^2} =  e^{-S_g[\bar{\boldsymbol{\phi}}_g]} \frac{\pi}{\sum_n m_n} \approx e^{-S_g[\bar{\boldsymbol{\phi}}_g]} \frac{\pi}{P_{\epsilon_g}-P_{\rm in}^{-1}},   \label{final-result}
\end{align}
where in the first line we have employed the eigenequation of $\mathcal{M}_g$, and in the subsequent line the Gaussian integral formula for complex variables has been applied. We reiterate that the integral computed in the above expression corresponds to the sum of $\mathcal{J}_i$ rather than $\mathcal{I}_i$ as discussed in Sec.~\ref{sec:qmandqft}, as the integral contour consists of all trajectories satisfying the gradient flow equation across the entire complex plane, i.e. the combination of different Lefschetz thimbles. In the final step, we took advantage of the expression for the sum of $m_n$, i.e. Eq.~\eqref{sum-of-mn}. 

We make some comments on the process and results of the calculation as follows:
\begin{itemize}
    \item It has been demonstrated that the saddle-point expansion method in false vacuum decay theory can be seamlessly applied to the path-integral approach to the LSS after some appropriate modifications and re-derivations. Although we have met the negative modes, and even complex modes of $S''_g[\bar{\boldsymbol{\phi}}_g]$, because all the eigenvalues of the gradient flow eigenequation Eq.~\eqref{full-eigenequation} are real and positive ($m_n \in \mathbb{R}, m_n>0$), the result is ensured to be real through the path integral, consistent with the argument as presented in Sec.~\ref{qmcase}. This aspect of the result is very crucial, as it directly relates to its physical interpretation as a probability.

    \item Although we have examined only one specific model, the rationale for applying the gradient flow equation to determine the steepest descent contours, as we discussed below, is general and can be applied to any kind of more complicated models. After this, we can compute arbitrary likelihoods of the form $\int \mathcal{D}\phi\,{\rm e}^{-S[\phi]}$ up to the second-order near the saddle points. This approach could be instrumental in providing more accurate predictions for the large-scale structure of the universe.
\end{itemize}
Next, we define the logarithm of the two likelihoods (matter and joint) as $\wp[\delta],\wp[\delta_g,\delta]$, thus we have
% --------------------------------------------------------------------------------------------------------------
\begin{align}
    \wp[\delta] &= {\rm ln} \,\mathcal{P}[\delta] = -{\rm ln}\,\left(\frac{\sqrt\pi}{D_1}\right)S[\bar {\boldsymbol{\phi}}] \, , \label{matterlog}
\end{align}
% --------------------------------------------------------------------------------------------------------------
\begin{align} 
    \wp[\delta_g,\delta]&= {\rm ln}\,\mathcal{P}[\delta_g,\delta] = -{\rm ln}\left(\frac{\pi}{\sum_{n} m_n}\right) S_g[\bar{\boldsymbol{\phi}}_g] \approx-{\rm ln}\left(\frac{\pi}{P_{\epsilon_g}-P_{\rm in}^{-1}}\right) S_g[\bar{\boldsymbol{\phi}}_g]\,.\label{jointlog}
\end{align}
Since neither the matrices $\mathcal{M}$ nor $\mathcal{M}_g$ depend on the overdensity fields $\delta$ and $\delta_g$, the eigenvalues will be field-independent. We can also see this in the expression for $m_n$ in Appendix~\ref{sec:eigenvalues}, where we provide a set of approximate solutions for them. Thus the logarithmic terms in Eqs.~\eqref{matterlog} and \eqref{jointlog} will be absorbed into the normalization factors. This absorption originates from the normalization factor introduced in the definition of the functional Dirac function Eq.~\eqref{functional-Dirac}. Consequently, this factor manifests in the expression for the likelihoods Eqs.~\eqref{matter-likelihood} and \eqref{joint-likelihood}.

We hereby provide a concise introduction to the renormalization scheme in this theory. Let us first consider the case in quantum field theory as an example. It appears that the application of saddle-point expansion may affect the renormalization, as the field decomposition introduces an additional fluctuation term $\varphi$, which needs a counterterm to cancel its UV dependence. However, in practice, if we aim to compute the radiative corrections to the path integral, we only need to directly use the renormalized effective action, rather than performing renormalization after the calculation of the saddle-point contributions. For example, some literature considers the 1PI \cite{Weinberg:1987vp, Andreassen:2016cvx} and 2PI \cite{Garbrecht:2015oea, Garbrecht:2015yza} effective actions\footnote{In the computation of the radiation corrections, it is often required that the effective action be convex to ensure physical consistency. A method to handle this issue using Green's function method can be found in \cite{Garbrecht:2015oea}.}. Returning to the present case, when there are field-independent terms in $\wp[\delta]$ and $\wp[\delta_g,\delta]$, they can be directly ignored due to the presence of the normalization factor $\mathcal{N}_{\delta^{(\infty)}}$. When these terms exhibit field-dependence, no additional counterterm needs to be introduced when incorporating loop corrections. Instead, the renormalized ``action" can be directly used.

Consequently, in this case, when defining the conditional likelihood of them, the terms appearing in the conditional likelihood will only include the saddle-point contributions of the two actions, which is
\begin{equation}
    \wp[\delta_g |\delta] = {\rm ln}\,\mathcal{P}[\delta_g|\delta] ={\rm ln}\, \frac{\mathcal{P}[\delta_g,\delta]}{\mathcal{P}[\delta]} = -S[\bar{\boldsymbol{\phi}}] +S_g[\bar{\boldsymbol{\phi}}_g] \, .
    \label{conditional-loglikelihood}
\end{equation}
Upon substituting the explicit expressions for $S[\bar{\boldsymbol{\phi}}]$ and $S_g[\bar{\boldsymbol{\phi}}_g]$, it becomes evident that our result is consistent with those presented in \cite{Cabass:2019lqx}. One notable observation is that the result following the saddle-point expansion appears to contribute only a factor that can be absorbed into the normalization factor, i.e. ${\rm e}^{-S[\bar{\phi}]}\int\mathcal{D}\varphi\, {\rm e}^{-\frac{1}{2}S''[\bar{\phi}]\varphi^2} \sim \mathcal{N} {\rm e}^{-S[\bar{\phi}]}$. However, this occurs solely because in this model $S''[\bar{\boldsymbol{\phi}}]$ does not exhibit dependence on $\delta_g$ and $\delta$. Suppose now we compute a more complex model, for instance, one where the action depends on $\boldsymbol{\phi}^3$. In this case the saddle-point solutions for the actions $\bar{\boldsymbol{\phi}}$ and $\bar{\boldsymbol{\phi}}_g$ will depend on $\delta_g$ and $\delta$, and $S''[\bar{\boldsymbol{\phi}}]$ will also exhibit dependence on these variables. In this way, the result of the saddle-point expansion introduces a factor that cannot be absorbed by normalization, and this factor will contribute to the final conditional likelihood. More complicated models will be left for future work to explore.

%====================================================================================================
\section{Summary and conclusion}
\label{sec:conclusion}

The path-integral approach to the EFTofLSS offers a novel method for encapsulating cosmological effects into a compact analytical framework, enabling the prediction of cosmological observables. In this work, we aim to apply the saddle-point expansion technique, traditionally employed in quantum field theory for calculating the decay rates, to this context to improve the precision of EFT likelihood calculations. As a result, after appropriately reformulating the likelihood expressions, this task reduces to computing the path integral over the field $\boldsymbol{\phi}$ in three-dimensional Euclidean space. Consequently, by applying the same reasoning as in quantum field theory, we can calculate the conditional likelihood, with the $\phi$ space replaced by the $\boldsymbol{\phi}$ space. 

A key challenge in the saddle-point expansion method is the treatment of the negative eigenvalues of the second derivative of the action at the saddle point. In this work, we present a systematic discussion of this aspect in three distinct cases. As an illustrative example, we compute the matter and joint likelihood under the assumption of a Gaussian distribution, retaining terms up to $J^2$ and $J_{g}^{2}$ order. For the matter likelihood, since there are no negative eigenvalues associated with $S''[\boldsymbol{\phi}]$, we can directly apply the Gaussian integral formula to obtain the result. However, when calculating the joint likelihood, the eigenvalues of $S''[\boldsymbol{\phi}_g]$ are often non-positive and even complex, necessitating an approximation of the original integral contour with a sum of the steepest descent contours in the field space. The treatment of these complex modes goes beyond the scope of the false vacuum decay theory discussion, primarily because the second derivative of the action in the path-integral approach to the LSS often takes a specific form, resulting in a general non-Hermitian three-dimensional matrix. Even so, we still demonstrate in Sec.~\ref{sec:example} that for this case, the Picard-Lefschetz theory remains applicable in this context by complexifying the action and using the gradient flow equation in the field space. We emphasize here again that although the integrals along each contour may yield complex results, their collective interference leads to a real final result.

Another point is that, as shown in Eqs.~\eqref{matter-answer} and \eqref{final-result}, the result of the saddle-point expansion is essentially equivalent to the saddle-point approximation, differing only by a multiplicative factor. In the example provided in Sec.~\ref{sec:example}, this factor is independent of $\delta_g$ and $\delta$, therefore it will be absorbed into the normalization constant and resulting in the same outcome as in \cite{Cabass:2019lqx}, as illustrated in Eq.~\eqref{conditional-loglikelihood}. However, this does not imply that the result of the saddle-point expansion is trivial, but rather that the action we have chosen in this example is truncated at $\boldsymbol{\phi}^2$ order. Furthermore, the method we present in this paper is general, and can be used to cope with more complicated situations.

Several unresolved issues remain in this work. For instance, we have not obtained an exact analytic solution of Eq.\eqref{hermite-equation}. The reason is that although the combination of the matrix $\mathcal{M}_g$ with its complex conjugate $\mathcal{M}_{g}^{*}$ is Hermitian, as illustrated in Eq.~\eqref{hermite-equation}, solving for this eigenequation is equivalent to solving a cubic equation, which does not have a simple analytic solution. Another important aspect concerns the treatment of higher-order terms. In the theory of false vacuum decay, the Lagrangian is typically truncated at the $\phi^4$ order due to the non-renormalizable nature of higher-order terms, which are strongly suppressed by the energy scale of new physics, $\mu$. This raises the question of whether a similar suppression applies to higher-order terms in our case. In \cite{Rubira:2024tea} and \cite{Rubira:2023vzw}, the authors computed the renormalization group equations for the partition function up to $J^2$ order and arbitrary $J^{n}$ order ($n \rightarrow \infty$). They found that, given the same initial conditions, the running of the bias parameter $b$ exhibits similar behavior. Does this imply when calculating the EFT likelihood, we can neglect the terms higher than $J^2$ (analogous to the non-renormalizable terms in quantum field theory)? In \cite{Rubira:2024tea} the authors perform a dimension analysis on the terms of all orders of $J$, however, do not draw an analogy to the case of quantum field theory. Then for the higher-order terms of $J$, are they also suppressed by the hard momentum cutoff $\Lambda$ introduced during the regularization process? e.g. consider the term $\int_{\boldsymbol{x}} J^{2}(\boldsymbol{x})O[\delta_{\Lambda}(\boldsymbol{x})]$, should we need to express this as $\frac{C_{*}}{\Lambda^{3+d_{O}}} \int_{\boldsymbol{x}} J^{2}(\boldsymbol{x})O[\delta_{\Lambda}(\boldsymbol{x})]$\footnote{Strictly speaking, this formulation is not precise. In quantum field theory, the energy scale of new physics $\mu$ and the momentum cutoff $\Lambda$ are two fundamentally distinct concepts \cite{Schwartz:2014sze}. The former is physical, introduced through dimensional regularization, and represents the energy scale at which we conduct our analysis. In contrast, the latter corresponds to the scale at which the theory breaks down in cutoff-based regularization schemes (for example, the Pauli-Villars regularization). It is often unphysical and can be taken to infinity in computations. We formulate this expression in this way to explicitly reflect the suppression of the operator.} such that this term will be suppressed by $\Lambda$, where $d_{O}$ is the dimension of the arbitrary operator $O$ and $C_{*}$ is the dimensionless ``coupling constant"? If similar reasoning as in quantum field theory can be applied, then these higher-order terms would contribute little to the partition function. Unfortunately, we have not found relevant discussions in any literature, so we refrain from drawing a definitive conclusion at this stage. 

Additionally, our calculation assumes a Gaussian initial condition. Extending our analysis to non-Gaussian initial conditions is essential, as it will facilitate a deeper investigation of the early universe. To address the challenge that SPT fails to adequately capture primordial non-Gaussianity, one possible solution is to construct sophisticated statistical quantities \cite{Wang:2024rdf}. Another approach is to employ functional methods similar to those presented in this work \cite{Nikolis:2024kbx, Vasudevan:2019ewf}. In the standard approach \cite{Assassi:2015fma, Schmidt:2010gw}, if the primordial probability distribution deviates from the Gaussian form, then the effective action will include an additional interaction term. This term introduces a cubic interaction in the field $\boldsymbol{\phi}$ when calculating the likelihood. The saddle-point expansion method would then require a careful treatment of this term. We leave a more detailed investigation of this issue to future work.

% -------------------------------------------------------------------------------------------------------------
\acknowledgments
JYK thanks Ying-Zhuo Li for his useful discussions. We thank the anonymous reviewer for his/her professional and valuable suggestions, which help us make our work complete and rigorous. We acknowledge the support by the National Science Foundation of China (No. 12147217, 12347163), the China Postdoctoral Science Foundation (No. 2024M761110), and the Natural Science Foundation of Jilin Province, China (No. 20180101228JC).

% --------------------------------------------------------------------------------------------------------------
\begin{appendix}
\section{The explicit cancellation of the imaginary part in Eq.~\eqref{zformula}}
\label{sec:cancellation}

In this appendix, we present the explicit calculation of the path integral given by Eq.~\eqref{zformula} and illustrate how the imaginary contributions from different $\mathcal{J}_i$ cancel out, ultimately yielding a real result.

In this example, what we need to calculate is
% ----------------------------------------------------------------------------------
\begin{equation}
    \mathcal{Z} = \sum_{i} \mathcal{J}_i = \mathcal{J}_{-1} + \mathcal{J}_0 +\mathcal{J}_1.
\end{equation}
After substituting the three paths, we can obtain
% ---------------------------------------------------------------------------------------------------------
\begin{align}
    \mathcal{Z} &= \exp\left(\frac{1}{4}\right)\int_{C_{-1}} {\rm d}z\, e^{-(z+1)^2} + \int_{C_0} {\rm d}z \, e^{\frac{1}{2}z^2} + \exp\left(\frac{1}{4}\right)\int_{C_{1}} {\rm d}z\, e^{-(z-1)^2} \nonumber \\
    &= \exp\left(\frac{1}{4}\right)\left(\int_{-\infty}^{0} {\rm d}z \, e^{-(z+1)^2} + \int_{0}^{-i\infty} {\rm d}z \, e^{-(z+1)^2}\right) + \int_{i \infty}^{-i\infty}{\rm d}z \,e^{\frac{1}{2}z^2} \nonumber\\ 
    &+\exp\left(\frac{1}{4}\right)\left(\int_{i\infty}^{0} {\rm d}z \, e^{-(z-1)^2} + \int_{0}^{\infty}{\rm d}z \, e^{-(z-1)^2}\right) \, , 
\end{align}
where we have already used $S''(-1) = S''(1) = 1/4$ and $S''(0) = 0$. We now examine the origin of the imaginary part of the integral. Firstly, the first term in the second line and the second term in the last line must be real, as both correspond to integrals along the real axis. At the same time, we can also obtain that
% -----------------------------------------------------------------------------------------------------------
\begin{equation}
    \int_{i \infty}^{-i\infty}{\rm d}z \,e^{\frac{1}{2}z^2} \xlongequal {u=iz} \int_{i \infty}^{-i \infty}{\rm d}(-iu)\, e^{-\frac{1}{2}u^2} =\int_{- \infty}^{ \infty}{\rm d}z \, e^{-\frac{1}{2}z^2}
\end{equation}
is real. Therefore, the only terms whose contributions are imaginary are the remaining two terms. This implies that the imaginary part of the path integral stems solely from the integration along the imaginary axis in this example. Meanwhile, it is straightforward to observe that
% ------------------------------------------------------------------------------------------------------------
\begin{align}
    \int_{0}^{-i\infty} {\rm d}z \, e^{-(z+1)^2} &\xlongequal{u=z+1} \int_{1}^{-i \infty} {\rm d}u \, e^{-u^2} \, ,\\ \int_{i\infty}^{0} {\rm d}z \, e^{-(z-1)^2} &\xlongequal{t=z-1} \int_{i\infty}^{-1} {\rm d}t \, e^{-t^2} \, .
\end{align}
A comparison of the two expressions above reveals that the integrands of the two integrals are identical, while the integration intervals reversed. Consequently, the contribution from both integrals cancel each other out, which is what we expect. Therefore, the overall result of the path integral is real
% -------------------------------------------------------------------------------------------------------------
\begin{align}
 \mathcal{Z} 
   & = \exp\left(\frac{1}{4}\right) \left(\int_{-\infty}^{1}{\rm d}z \, e^{-z^2} + \int_{-1}^{\infty}{\rm d}z \, e^{-z^2}\right) + \int_{-\infty}^{\infty}{\rm d}z \, e^{-\frac{1}{2}z^2} \nonumber \\
   & = \exp\left(\frac{1}{4}\right) \left(\int_{-\infty}^{\infty}{\rm d}z \, e^{-z^2} +\int_{-1}^{1}{\rm d}z \, e^{-z^2}\right) + \sqrt{2\pi} \nonumber \\
   & \approx \exp\left(\frac{1}{4}\right) \left(\sqrt{\pi} + 0.75\right)+\sqrt{2\pi}\,,
\end{align}
where, from the second line to the third line, we have substituted the approximate value of the Gaussian error function.

On the other hand, if we use the approximate formula for $\mathcal{I}_i$ given in Eq.~\eqref{approxma} rather than the exact the integral $\mathcal{J}_i$, we will obtain
% -------------------------------------------------------------------------------------------------------------
\begin{equation}
      \mathcal{I}_{-1} = \sqrt{\pi} \, {\rm exp}\left(\frac{1}{4}\right)\,, \,\,\, \mathcal{I}_0 = \sqrt{-2\pi}\,, \,\,\, \mathcal{I}_{-1} = \sqrt{\pi}\, {\rm exp}\left(\frac{1}{4}\right) \,.
\end{equation}
We can observe that in this case $\mathcal{I}_{-1}$ and $\mathcal{I}_{1}$ have no imaginary part, whereas $\mathcal{I}_0$ exhibits the opposite behavior. Let us explain why the approximation $\mathcal{J}_i \sim \mathcal{I}_i$ fails to result in the necessary cancellation of the imaginary parts. When we approximate the original integral path by the sum of different Lefschetz thimbles, the reality and convergence of the entire integral are ensured by the Picard-Lefschetz theory \cite{Andreassen:2016cvx}. However, when we replace $\mathcal{J}_i$ with $\mathcal{I}_i$, each $\mathcal{I}_i$ captures only the contribution from a small region around the corresponding saddle point, rather than the contribution from the entire Lefschetz thimble. This become evident from the explicit formulation of $\mathcal{I}_i$,
% ------------------------------------------------------------------------------------------------------------
\begin{equation}
   \mathcal{I}_i= \sqrt{\frac{2\pi}{S''_{E}[s_i]}}\, e^{-S_{E}[s_i]} \,, 
\end{equation}
which only encapsulates the local contributions in the immediate vicinity of the saddle point. This is analogous to the situation in the WKB approximation \cite{Ai:2019fri, Andreassen:2016cvx} in quantum mechanics, where we use the classical trajectory to approximate the original one. We can check this in the context of this example: for $\mathcal{I}_{-1/1}$ associated with $s_{-1/1}$, since the integral along the path $C_{-1/1}$ around the saddle point takes along the real axis, its contribution is real, as shown in Eq.~\eqref{real-answer}. Similarly, the integral along the path $C_{0}$ around $z=0$ is taken along the imaginary axis, hence the contribution of $\mathcal{I}_0$ is imaginary. We emphasize that in this work, when applying the saddle-point expansion method to evaluate the likelihood, we will use $\mathcal{J}_i$ rather than $I_i$ as our expression, as illustrated in Sec.~\ref{sec:example}.

% ------------------------------------------------------------------------------------------------------------
\section{The contribution of the saddle-point}
\label{sec:saddle-point-contribution}

To complete our work, here we provide a brief review of the calculation of the saddle-point contributions, as proposed in \cite{Cabass:2019lqx}. We begin by considering the equations of motion
\begin{align}
    \frac{\delta S[\boldsymbol{\phi}]}{\delta\boldsymbol{\phi}}\bigg|_{\boldsymbol{\phi}  =\bar{\boldsymbol{\phi}}} &= \mathcal{J}^a + \mathcal{M}^{ab} \bar{\boldsymbol{\phi}}^{b}  =0\,,  \\
      \frac{\delta S_g[\boldsymbol{\phi}_g]}{\delta\boldsymbol{\phi}_g}\bigg|_{\boldsymbol{\phi}_g  =\bar{\boldsymbol{\phi}}_g} &= \mathcal{J}^{a}_{g} + \mathcal{M}^{ab}_{g} \bar{\boldsymbol{\phi}}^{b}_{g}  =0\,.  
\end{align}
These two equations can be readily solved once the forms of $\mathcal{M}$ and $\mathcal{M}_g$ are known, i.e. through Eqs.~\eqref{m-form} and \eqref{mg-form}. The solution for the former case is
\begin{align}
    \bar{\boldsymbol{\phi}}^{a} = -(\mathcal{M}^{ab})^{-1}\mathcal{J}^{b} = - 
    \begin{pmatrix}
        \frac{P_{\rm in}^{-1}}{D_{1}^{2}}  & \frac{1}{iD_{1}} \\ \frac{1}{iD_1} & 0
    \end{pmatrix}
    \begin{pmatrix}
        i \delta \\ 0
    \end{pmatrix} = 
    \begin{pmatrix}
        \frac{{i \delta}}{D_{1}^{2} P_{\rm in}}\\ 
        \frac{\delta}{D_1}
    \end{pmatrix}. 
    \label{bar-phi}
\end{align}
and for the latter case
\begin{align}
    \bar{\boldsymbol{\phi}}_g^{a} & = -(\mathcal{M}^{ab}_{g})^{-1}\mathcal{J}^{b}_{g} =\frac{1}{-2b_1D_{1}^{2}P_{\epsilon_g \epsilon_m}+P_{\epsilon_g}D_{1}^{2}-P_{\epsilon_g \epsilon_m}P_{\rm in}^{-1} } \times \nonumber\\ &\begin{pmatrix}
        D_{1}^{2} &  -b_1D_{1}^{2}-P_{\epsilon_g \epsilon_m}P_{\rm in}^{-1} & iD_1P_{\epsilon_g \epsilon_m} \\ -b_1D_{1}^{2}-P_{\epsilon_g \epsilon_m}P_{\rm in}^{-1} &b_{1}^{2}D_{1}^{2}+P_{\epsilon_g}P_{\rm in}^{-1} & ib_1D_1P_{\epsilon_g \epsilon_m}-iD_1P_{\epsilon_g} \\ iD_1P_{\epsilon_g \epsilon_m} & ib_1D_1P_{\epsilon_g \epsilon_m}-i D_1P_{\epsilon_g} & P_{\epsilon_g \epsilon_m}^{2} 
    \end{pmatrix}
    \begin{pmatrix}
        i\delta_g \\ i\delta \\ 0 
    \end{pmatrix} \nonumber\\
    & \approx \frac{1+2b_1P_{\epsilon_g \epsilon_m}/P_{\epsilon_g}}{D_{1}^{2}P_{\epsilon_g}} \begin{pmatrix}
        iD_{1}^{2}\delta_g -ib_1D_{1}^{2}\delta - iP_{\epsilon_g \epsilon_m}P_{\rm in}^{-1} \delta \\ i\delta_g(-b_1 D_{1}^{2}-P_{\epsilon_g \epsilon_m}P_{\rm in}^{-1}) + i\delta(D_{1}^{2}b_{1}^{2} + P_{\epsilon_g}P_{\rm in}^{-1}) \\
        -D_{1}P_{\epsilon_g \epsilon_m}\delta_g + i\delta(ib_1D_1P_{\epsilon_g \epsilon_m}-iD_1P_{\epsilon_g})
    \end{pmatrix}\, \,,
\end{align}
where from the second step to the third step we have assumed $P_{\epsilon_g \epsilon_m} \ll P_{\epsilon_g}$ and used the Taylor expansion $(1-x)^{-1} \sim 1+x$. For convenience, we define $X_g = \frac{i(\delta_g-b_1\delta)}{P_{\epsilon_g}}$, which is the solution of the equation of motion for $X_g$ when $P_{\epsilon_g\epsilon_m} = 0$. Then the expression for 
$\boldsymbol{\phi}_g$ can be written as
\begin{equation}
    \bar{\boldsymbol{\phi}}_g = \begin{pmatrix}
        X_g + \frac{P_{\epsilon_g \epsilon_m}}{P_{\epsilon_g}}(2b_{1}X_{g} - \frac{i\delta}{D_{1}^{2}P_{\rm in}}) \\
        \frac{i\delta}{D_{1}^{2}P_{\rm in}} - b_1X_g -\frac{b_1P_{\epsilon_g \epsilon_m}}{P_{\epsilon_g}}(2b_{1}X_{g} - \frac{i\delta}{D_{1}^{2}P_{\rm in}}) -\frac{P_{\epsilon_g \epsilon_m}X_g}{D_{1}^{2}P_{\rm in}} \\
        \frac{\delta}{D_1} + \frac{iP_{\epsilon_g \epsilon_m}X_g}{D_1} 
    \end{pmatrix}\,. \label{bar-phig}
\end{equation}
After obtaining the expressions in Eqs.~\eqref{bar-phi} and \eqref{bar-phig}, we can substitute them into the actions Eqs.~\eqref{matter-action} and \eqref{joint-action} to calculate the contributions of the saddle points to the conditional likelihood. This has already been accomplished in \cite{Cabass:2019lqx}, and we enumerate here all terms up to the third order in both $\delta$ and $\delta_{g}$.

We extract only the terms that are field-dependent. At quadratic order we have
\begin{equation}
    \begin{split}
(S[\bar{\boldsymbol{\phi}}]-S[\bar{\boldsymbol{\phi}}_g])^{(2)} = 
{-\frac{1}{2}}\int_{\boldsymbol{k}}\frac{|{\delta_g(\boldsymbol{k})-\delta_{g,{\rm det}}[\delta]
(\boldsymbol{k})}|^2}{P_{\epsilon_g}(k) - 2b_1P_{\epsilon_g\epsilon_m}(k)} + \Delta\wp\,\,.
\end{split}
\end{equation}
in which we have defined $\delta_{g,\rm det}[\delta_{\rm fwd}[\delta_{\rm in}]] = \delta_{g, \rm fwd}[\delta_{\rm in}]$. And
\begin{equation}
    \Delta\wp= \int_{\boldsymbol{k}}\frac{P_{\epsilon_g\epsilon_m}(k)
\big(\delta_g(\boldsymbol{k})-\delta_{g,{\rm det}}^{(1)}[\delta](\boldsymbol{k})\big)\delta(-\boldsymbol{k})}{P_{\epsilon_g}(k)P_{\rm L}(k)}\,\,.
\end{equation}
In which we have defined $P_{L}(k) =D_{1}^{2}P_{\rm in}(k) $. At cubic order there are three terms that contributed to the conditional likelihood, they are 
\begin{equation}
  (S[\bar{\boldsymbol{\phi}}]-S[\bar{\boldsymbol{\phi}}_g])^{(3)} \supset  2\int_{\boldsymbol{k}}\frac{b_1P_{\epsilon_g\epsilon_m}(k)}{P_{\epsilon_g}(k)} 
\frac{\big(\delta_g(\boldsymbol{k})-b_1\delta(\boldsymbol{k})\big)\,
\delta^{(2)}_{g,{\rm det}}[\delta](-\boldsymbol{k})}{P_{\epsilon_g}(k)}\,\,,
\end{equation}
and
\begin{equation}
  (S[\bar{\boldsymbol{\phi}}]-S[\bar{\boldsymbol{\phi}}_g])^{(3)} \supset  {-{}}\int_{\boldsymbol{k}}
\frac{P_{\epsilon_g\epsilon_m}(k)}{P_{\epsilon_g}(k)}\frac{\delta(\boldsymbol{k})\delta^{(2)}_{g,{\rm det}}[\delta](-\boldsymbol{k})}{P_{\rm L}(k)}\,\,.
\end{equation}
Finally, we have
\begin{align}
    (S[\bar{\boldsymbol{\phi}}]-S[\bar{\boldsymbol{\phi}}_g])^{(3)} &\supset {-{}}\int_{\boldsymbol{k}}\frac{\delta_g(\boldsymbol{k})-\delta_{g,{\rm det}}^{(1)}(\boldsymbol{k})}{P_{\epsilon_g}(k)} 
\int_{\boldsymbol{p}_1,\boldsymbol{p}_2}\bigg[(2\pi)^3\delta^{(3)}(-\boldsymbol{k}-\boldsymbol{p}_{12})\,K_{g,{\rm det},2}(-\boldsymbol{k};\boldsymbol{p}_1,\boldsymbol{p}_2) \nonumber \\
&\times\bigg(\frac{P_{\epsilon_g\epsilon_m}(p_2)}{P_{\epsilon_g}(p_2)}
\delta(\boldsymbol{p}_1)\big(\delta_g(\boldsymbol{p}_2) - \delta_{g,{\rm det}}^{(1)}(\boldsymbol{p}_2)\big)\bigg)  (\boldsymbol{p}_1\to\boldsymbol{p}_2)\bigg]\,\,.
\end{align}

\section{The eigenvalues of $\mathcal{M}_g$}
\label{sec:eigenvalues}
In this appendix, we will provide details regarding the computation of the eigenvalues of the matrix $\mathcal{M}_g$. We begin with the Hermitian eigenequation constructed from $\mathcal{M}_g$
\begin{equation}
     \begin{pmatrix}
         \boldsymbol{0}  & \mathcal{M}^{*}_{g} \\
         \mathcal{M}_{g} & \boldsymbol{0}
        \end{pmatrix}
    \begin{pmatrix}
        \chi_{n}(\boldsymbol{k}) \\ {\chi}_{n}^{*}(\boldsymbol{k})
    \end{pmatrix}
     =  m_{n}\begin{pmatrix}
         \chi_{n}(\boldsymbol{k}) \\ {\chi}_{n}^{*}(\boldsymbol{k})
     \end{pmatrix} \, .
\end{equation}
For simplicity, in the context of our analysis, we will perform a series of substitutions on the parameters within the matrix, writing 
\begin{align}
    P_{\epsilon_g} \equiv a \,,\, P_{\epsilon_g\epsilon_m} \equiv b\, ,\, b_1D_1 \equiv c\,,\, D_1\equiv d\,,\, P_{\rm in}^{-1} \equiv e\,.\,
\end{align}
then the eigenequation can be expressed as
\begin{equation}
    \begin{pmatrix}
        0 & 0& 0 &a&b&-ic \\
        0&0&0&b&0& -id \\
        0&0&0&-ic &-id &e\\
        a&b&ic&0&0&0\\
        b&0&id&0&0&0\\       ic &id&e&0&0&0        
    \end{pmatrix}
      \begin{pmatrix}
        \chi_{n}(\boldsymbol{k}) \\ {\chi}_{n}^{*}(\boldsymbol{k})
    \end{pmatrix}
     =  m_{n}\begin{pmatrix}
         \chi_{n}(\boldsymbol{k}) \\ {\chi}_{n}^{*}(\boldsymbol{k})
     \end{pmatrix} \, .
\end{equation}
There are six eigenvalues associated with this equation, of which we only need to extract the three distinct solutions (corresponding to the eigenvalues of $\mathcal{M}_g$). The expressions of them are (after some analytical approximations)
\begin{align}
    m_1 & = \frac{a-e}{3} -\frac{c_1}{3\left(\frac{1}{2}(c_2 +\sqrt{4c_{2}^{2}-4c_{1}^{3}})\right)^{\frac{1}{3}}}-\frac{\left(\frac{1}{2}(c_2+\sqrt{4c_{2}^{2}-4c_{1}^{3}})\right)^{\frac{1}{3}}}{3} \,; \\
    m_2 &= \frac{a-e}{3} -\frac{c_1}{3\left(-\frac{1}{2}+\frac{\sqrt{3}i}{2}\right)\left(\frac{1}{2}(c_2 +\sqrt{4c_{2}^{2}-4c_{1}^{3}})\right)^{\frac{1}{3}}} - \frac{(-\frac{1}{2}+\frac{\sqrt{3}i}{2})\left(\frac{1}{2}(c_2+\sqrt{4c_{2}^{2}-4c_{1}^{3}})\right)^{\frac{1}{3}}}{3} \,; \\
    m_3 &= \frac{a-e}{3} -\frac{c_1}{3\left(-\frac{1}{2}-\frac{\sqrt{3}i}{2}\right)\left(\frac{1}{2}(c_2 +\sqrt{4c_{2}^{2}-4c_{1}^{3})}\right)^{\frac{1}{3}}} - \frac{(-\frac{1}{2}-\frac{\sqrt{3}i}{2})\left(\frac{1}{2}(c_2+\sqrt{4c_{2}^{2}-4c_{1}^{3}})\right)^{\frac{1}{3}}}{3}\,,
\end{align}
where we have defined 
\begin{align}
    c_1 &= 3ae +3b^2 +3c^3+3d^2+(-a+e)^3 \, , \\
    c_2 &= \frac{27ad^2}{2}- \frac{27b^2e}{2} -27bcd +\frac{9(e-a)(ae+b^2+c^2+d^2)}{2} +(e-a)^3 \,.
\end{align}
\end{appendix}
Note that in the expressions of $m_2$ and $m_3$, there is the existence of the imaginary unit $i$. However, we emphasize that this is the consequence of analytical approximations that must be made when solving cubic equations. Since we are solving the eigenequation of a Hermitian operator, all exact eigenvalues must, by definition, be real. Even so, we can still check that the sum of these three eigenvalues is real
\begin{equation}
    m_1+m_2+m_3 = a-e = P_{\epsilon_g} - P_{\rm in}^{-1} \,. \label{sum-of-mn}
\end{equation}
We emphasize here again this is just the approximate solution, and will use the above expression as the result of $\sum_n m_n$ in Eq.~\eqref{final-result}.

%%%%%%%%%%%%%%%%%%%%%%%%%%%%%%%%%%%%%%%%%%%%%%%%%%
\bibliographystyle{JHEP}
\bibliography{reference}

\end{document}